\documentclass{aa}

\usepackage{txfonts}

\DeclareRobustCommand{\VAN}[3]{#2}
\let\VANthebibliography\thebibliography
\def\thebibliography{\DeclareRobustCommand{\VAN}[3]{##3}\VANthebibliography}

\usepackage{graphicx}	
\usepackage{amssymb}	

\usepackage{xcolor}
\usepackage{overpic}
\usepackage{subfigure}
\usepackage{booktabs}
\usepackage{multirow}
\usepackage{caption}
\usepackage[flushleft]{threeparttable}

\usepackage[version=4]{mhchem}

\usepackage[colorlinks=true,allcolors=blue,urlcolor=blue]{hyperref}

\newcommand{\freeeos}{\texttt{FreeEOS}}
\newcommand{\blue}{\texttt{Blue}}
\newcommand{\stagger}{\texttt{Stagger}}
\newcommand{\balder}{\texttt{Balder}}
\newcommand{\multitd}{\texttt{Multi3D}}
\newcommand{\mean}[1]{\ensuremath{\langle #1 \rangle}}

\begin{document} 

\title{3D Stagger model atmospheres with FreeEOS}
\subtitle{I. Exploring the impact of microphysics on the Sun}
\titlerunning{3D \stagger{} model atmospheres with \freeeos{}}

\author{Yixiao Zhou\inst{\ref{sac}}
\and
Anish M. Amarsi\inst{\ref{uu1}}
\and
Victor Aguirre B{\o}rsen-Koch\inst{\ref{dark}}
\and
Klara G. Karlsmose\inst{\ref{sac}}
\and
Remo Collet\inst{\ref{sac}}
\and
Thomas Nordlander\inst{\ref{anu},\ref{astro3d}}
}

\institute{
\label{sac}Stellar Astrophysics Centre, Department of Physics and Astronomy, Aarhus University, Ny Munkegade 120, DK-8000 Aarhus C, Denmark\\
\email{yixiao.zhou@qq.com}
\and
\label{uu1}Theoretical Astrophysics, Department of Physics and Astronomy, Uppsala University, Box 516, SE-751 20 Uppsala, Sweden\\
\email{anish.amarsi@physics.uu.se} 
\and
\label{dark}DARK, Niels Bohr Institute, University of Copenhagen, Jagtvej 128, 2200 Copenhagen, Denmark
\and
\label{anu}Research School of Astronomy and Astrophysics, Australian National University, ACT 2611, Canberra, Australia
\and
\label{astro3d}ARC Centre of Excellence for All Sky Astrophysics in 3 Dimensions (ASTRO 3D), Australia
}

\abstract{
  Three-dimensional radiation-hydrodynamics (3D RHD) simulations of stellar surface convection provide valuable insights into many problems in solar and stellar physics. 
  However, almost all 3D near-surface convection simulations to date are based on solar-scaled chemical compositions, which limit their application on stars with peculiar abundance patterns. 
  To overcome this difficulty, we implement the robust and widely-used \freeeos{} equation of state and our \blue{} opacity package into the \stagger{} 3D radiation-magnetohydrodynamics code. We present a new 3D RHD model of the solar atmosphere, and demonstrate that the mean stratification as well as the distributions of key physical quantities are in good agreement with those of the latest \stagger{} solar model atmosphere. 
  The new model is further validated by comparing against solar observations. The new model atmospheres reproduce the observed flux spectrum, continuum centre-to-limb variation, and hydrogen line profiles at a satisfactory level, thereby confirming the realism of the model and the underlying input physics. 
  These implementations open the prospect for studying other stars with different $\alpha$-element abundance, carbon-enhanced metal-poor stars and population II stars with peculiar chemical compositions using 3D \stagger{} model atmospheres.
}


\date{Received / Accepted}

\maketitle



\section{Introduction}

  Stellar atmosphere models are indispensable tools for the quantitative interpretation of astronomical observations. For late-type stars, although the majority of theoretical atmosphere models are computed assuming one-dimensional (1D) geometry, hydrostatic equilibrium and phenomenological theories of convection such as the mixing length theory (MLT, \citealt{1958ZA.....46..108B}), the use of three-dimensional radiation-hydrodynamics (3D RHD) stellar atmosphere models are thriving in recent years, partly driven by more detailed observational data and the rapid growth in computing power. 
  In 3D RHD models, sometimes referred to as near-surface convection simulations, the motion of fluid is computed from first principles by solving the equation of mass and momentum conservation as well as the energy conservation equation coupled with the equation of radiative transfer in 3D space for each timestep. Although the current study ignores the effects of magnetic fields,
  they can be included in the models by adding the induction equation, Amp\`ere's circuital law and Ohm's law to the equation system. 

  The 3D RHD models have proven to be superior to their 1D counterparts in all aspects and shed light on many problems in stellar physics. The early simulations by, for example, \citet{1985SoPh..100..209N} and \citet{1989ApJ...342L..95S} provided valuable insight into how convection operates in the near-surface layers of low-mass stars: Rather than distinct and coherent fluid parcels assumed in the MLT, convective regions show finger-like downflows that
  merging together as they descend from the photosphere before finally reaching the bottom of the simulation domain.
  Conservation of mass forces the relatively hot material to rise back up through the thin optical surface, forming so-called granules. Detailed solar simulations presented in \citet{1998ApJ...499..914S} further confirm this picture, and their work demonstrated excellent agreement between simulation and observation in the granulation pattern.
  
  These \textit{ab initio} simulations have enabled the prediction of various observables in a parameter-free manner. 
  The most remarkable breakthrough brought by 3D RHD models is associated with spectral line profiles: Predicted spectral lines broadening, blueshifts and bisectors agree excellently with observations \citep{2000A&A...359..729A,2013A&A...554A.118P}, to a degree that cannot be achieved by 1D models even if free parameters in the latter can be adjusted to fit the measured line profiles. This renders 3D model atmospheres a powerful tool for elemental abundance determinations and leads to a revision of the standard solar chemical composition \citep{2005ASPC..336...25A,2009ARA&A..47..481A,2011SoPh..268..255C,2021A&A...653A.141A}.
  Moreover, 3D RHD models perform well in the case of centre-to-limb variations of intensity \citep{2013A&A...554A.118P,2017A&A...597A.137K}, making them useful for deriving limb-darkened stellar radii and effective temperatures from interferometric measurements \citep{2018MNRAS.477.4403W,2020A&A...640A..25K,2020MNRAS.493.2377R}.
  3D simulations of stellar surface convection have also contributed to the field of helioseismology. \citet{1999A&A...351..689R} showed that the discrepancy between theoretical and measured solar pressure mode frequencies can be reduced by combining the averaged 3D model with 1D interior model then computing the theoretical oscillation frequency based on such a patched model (see also \citealt{2016A&A...592A.159B} and \citealt{2017MNRAS.464L.124H}). The reason for the better agreement is that the convective turbulence is self-consistently described in 3D simulations, thereby resulting in more realistic pressure stratification in the near-surface convective layers.

  3D hydrodynamical simulations of the solar near-surface convective region and solar atmosphere have been carried out by several research groups with independent codes, among others, \texttt{ANTARES} \citep{2010NewA...15..460M}, \texttt{Bifrost} \citep{2011A&A...531A.154G}, \texttt{CO$^5$BOLD} \citep{2012JCoPh.231..919F}, \texttt{MURaM} \citep{2005A&A...429..335V} and \stagger{} \citep{1995...Staggercodepaper}. The \texttt{Bifrost} solar simulation extends from the near-surface convection zone upwards to the corona to investigate processes in the transition region and the solar chromosphere \citep{2016A&A...585A...4C}. The magneto-hydrodynamical simulations of the Sun constructed with the \texttt{MURaM} code have provided valuable insights into our understanding of sunspots \citep{2009Sci...325..171R} and the solar small-scale dynamo \citep{2007A&A...465L..43V}. Reference solar simulations computed with the \texttt{CO$^5$BOLD}, \texttt{MURaM} and \stagger{} code were compared in detail by \citet{2012A&A...539A.121B}, who found good quantitative agreement between the three models.
  
  Meanwhile, near-surface convection simulations with \texttt{CO$^5$BOLD}, \texttt{MURaM} and \stagger{} were constructed for a variety of stars including warm turn-off stars \citep{2002ApJ...567..544A}, M-type dwarfs \citep{2006A&A...459..599L}, metal-poor benchmark stars \citep{2009PASA...26..330C,2018MNRAS.475.3369C} and extremely metal-poor stars \citep{2006ApJ...644L.121C,2023A&A...672A..90L}. Also, grids of 3D model atmospheres, such as the \texttt{CIFIST} grid \citep{2009MmSAI..80..711L,2022A&A...661A..76B}, the \citet{2013ApJ...769...18T} grid, and the \stagger{} grid (\citealt{2013A&A...557A..26M}; Rodr{\'\i}guez D{\'\i}az et al., in preparation), are now available, covering a large area of the Hertzsprung-Russell diagram and spanning a wide metallicity range. 
    
  However, most of the 3D model atmospheres to date are constructed based on solar-scaled chemical compositions and a fixed value of abundance enhancement of $\alpha$ elements\footnote{In the \stagger{} grid, the abundance of $\alpha$ process elements O, Ne, Mg, Si, S, Ar and Ca are enhanced by 0.4 dex, $\rm [\alpha / Fe] = 0.4$, for metal-poor models with $\rm [Fe/H] \le -1$. The notation ${\rm [A/B]} = \log(n_{\rm A} / n_{\rm B}) - \log(n_{\rm A} / n_{\rm B})_{\odot}$, where $n_{\rm A} / n_{\rm B}$ and $(n_{\rm A} / n_{\rm B})_{\odot}$ represent number density ratio between element A and B in the star and the Sun, respectively.}. Although this is usually an acceptable approximation for solar-type stars (i.e.~F- and G-type dwarfs), the validity of the model atmosphere is in doubt when applied to, for example, relatively metal-rich ($\rm [Fe/H] \sim -0.5$) halo stars with high $\alpha$ element abundance ($\rm [\alpha / Fe]$ up to $\sim 0.35$, see \citealt{2010A&A...511L..10N}) because $\alpha$-enhancement is not considered at this metallicity. 
  Neglecting variations in the abundance of individual elements may result in more significant systematic errors when investigating stars with peculiar abundance patterns such as carbon-enhanced extremely metal-poor stars, whose carbon and oxygen abundance is usually enhanced by at least 1 dex with respect to solar-scaled values \citep{2015ARA&A..53..631F}. 
  
  Therefore, model atmospheres with chemical composition tailored for individual cases are necessary for the aforementioned stars, given their importance in revealing the chemical evolution history of the Milky Way and the early Universe. \citet{2017A&A...598L..10G} and \citet{2018IAUS..334..364S} have made pioneering efforts in this direction by generating 3D model atmospheres for carbon-enhanced metal-poor stars.
  To allow realistic \stagger{} simulations with arbitrary chemical composition we utilise an open-source EOS code and an opacity package developed in-house (Sect.~\ref{sec:physics}), which form the basis of our new models.
  As the input physics has changed, the resulting model atmospheres need to be validated before any scientific application. The Sun is a natural test bench for all theoretical stellar models due to the rich observational constraints available. Therefore, in this work, we validate the newly implemented input physics and the resulting 3D solar model atmosphere by comparisons with previously published results in terms of mean structure and horizontal distribution of key quantities (Sect.~\ref{sec:solar-model}), and by comparing model-predicted quantities with corresponding solar observations (Sect.~\ref{sec:compare-obs}).

\section{Input physics} \label{sec:physics}

\subsection{Equation of state} \label{sec:feos}

\begin{table}
\centering
\caption{List of elements included in the \freeeos{} calculation and the corresponding solar abundance adopted in this work. The abundance of element ``X'' is $\epsilon_{\rm X} = \log(n_{\rm X} / n_{\rm H}) + 12$ with $n_{\rm H}$ being the hydrogen number density.
\label{tb:abundance}}
{\begin{tabular*}{\columnwidth}{@{\extracolsep{\fill}}cccc}
\toprule[2pt]
  Atomic number & Element & $\epsilon_{\rm AGSS09}$ & $\epsilon_{\rm AAG21}$
  \\
\midrule[1pt]
  1  & H  & 12.00 & 12.00
  \\ 
  2  & He & 10.93 & 10.914
  \\ 
  6  & C  & 8.43  & 8.46
  \\  
  7  & N  & 7.83  & 7.83
  \\ 
  8  & O  & 8.69  & 8.69
  \\
  10 & Ne & 7.93  & 8.06
  \\
  11 & Na & 6.24  & 6.22
  \\
  12 & Mg & 7.60  & 7.55
  \\
  13 & Al & 6.45  & 6.43
  \\
  14 & Si & 7.51  & 7.51
  \\
  15 & P  & 5.41  & 5.41
  \\
  16 & S  & 7.12  & 7.12
  \\
  17 & Cl & 5.50  & 5.31
  \\
  18 & Ar & 6.40  & 6.38
  \\
  20 & Ca & 6.34  & 6.30
  \\
  22 & Ti & 4.95  & 4.97
  \\
  24 & Cr & 5.64  & 5.62
  \\
  25 & Mn & 5.43  & 5.42
  \\
  26 & Fe & 7.50  & 7.46
  \\
  28 & Ni & 6.22  & 6.20
  \\
\bottomrule[2pt]
\end{tabular*}}
\end{table}

  \freeeos{} is an open-source EOS code by \citet{2004Irwin...feos1,2012ascl.soft11002I} for stellar interior and atmosphere conditions. The EOS covers a wide temperature and density range that blankets both the lower atmosphere and the core of low-mass stars. It is widely adopted in stellar evolution codes, such as \texttt{GARSTEC} \citep{2008Ap&SS.316...99W} and \texttt{MESA} \citep{2023ApJS..265...15J}, and recently applied to \texttt{MURaM} magneto-hydrodynamical simulations to study small-scale dynamo in cool stars with different spectral types and metallicities \citep{2022A&A...663A.166B,2023A&A...669A.157W}.
  The EOS includes 20 elements (Table \ref{tb:abundance}) as well as $\rm H_2$ and $\rm H_2^+$ molecules. At each density and temperature (or pressure) pair, chemical equilibrium is computed by adjusting the number density of various species (atoms, ions and electrons) such that the Helmholtz free energy is minimised, with the charge-neutral condition and particle number conservation (also known as abundance conservation) as constraints. The free energy minimisation technique is often used in EOS calculations (e.g.~\citealt{1988ApJ...331..794H}) because when temperature and volume are fixed, the Helmholtz free energy is a minimum at equilibrium.
  For detailed explanations of the code, we refer the readers to the \freeeos{} documentations \citep{2004Irwin...feos1,2004Irwin...feos2}\footnote{Available at \url{http://freeeos.sourceforge.net/documentation.html}}.

  We used the most realistic free energy model implemented in the \freeeos{} (named EOS1) in our calculation. The EOS1 option takes into account all ionisation stages of the 20 included elements, arbitrarily relativistic and degenerate free electrons, higher order Coulomb effects through a Coulomb factor on the first-order Debye-H{\"u}ckel term, and an occupation probability formulation (similar to the \citealt{1988ApJ...331..815M} EOS, hereafter MHD EOS) for pressure ionisation.
  It yields good agreement with the OPAL EOS \citep{2002ApJ...576.1064R} by design. Tests performed by \citet{2004Irwin...feos2} show the thermodynamical quantities predicted by the \freeeos{} EOS1 option differ from that of the OPAL EOS by less than 0.2\% for solar conditions.
  
  We adopted two solar abundance mixtures in this work, the \citet[hereafter AGSS09]{2009ARA&A..47..481A} and \citet[AAG21]{2021A&A...653A.141A} solar abundance. The input abundances for 20 elements included in the EOS are listed in Table \ref{tb:abundance}. 
  Thermal (gas plus radiation) pressure given by the two sets of abundances differs by about 0.7\% in the mass density and temperature area of interest ($-10 < \log(\rho / {\rm [g\,cm^{-3}]}) < -4$; $3.5 < \log(T / {\rm [K]}) < 4.5$, where $\rho$ and $T$ are mass density and temperature, respectively). 
  The difference is almost due entirely to different helium abundance adopted in AGSS09 and AAG21. The effect of metals on thermal pressure is negligible.
  As detailed in Appendix \ref{sec:compare-EOS}, key quantities from the \freeeos{} with the AGSS09 solar abundance are compared with a modified version of the MHD EOS \citep[the EOS employed in previous \stagger{} simulations]{2013ApJ...769...18T}. We found good agreement between the two EOSs, which further validates our \freeeos{} results.

  With the aforementioned setup and inputs, we generated \freeeos{} tables in the format required by the \stagger{} code. Our EOS tables are based on mass density and internal energy per mass $e_m$ as independent variables, as internal energy per mass rather than temperature is the fundamental variable in the \stagger{} code (see Sect.~2 of \citealt{1995...Staggercodepaper}). Densities and internal energies are equidistant in logarithm space. Density $\log\rho$ ranges from $-13.9$ to 0 $\rm \log[g\,cm^{-3}]$ in steps of $0.05$ while $\log e_m$ ranges from $11$ to $14$ $\rm \log[erg\,g^{-1}]$ in steps of $0.005$. The resolution of our EOS table is higher than that previously used in the \stagger{} code. 
  
  The EOS table stores thermal pressure $P_{\rm ther}$, temperature $T$ and electron number density $n_e$ and their partial derivatives with respect to two independent variables, i.e.~$(\partial\ln f / \partial\ln\rho)_{e_m}$ and $(\partial\ln f / \partial\ln e_m)_{\rho}$ where $f \in \{ P_{\rm ther}, T, n_e \}$. Note that these thermodynamic derivatives are obtained directly from EOS calculations and the Maxwell relations, therefore no interpolation is needed.

\subsection{Opacities}

\subsubsection{The \blue{} opacity code}\label{sec:op-blue}

\begin{table}
\begin{threeparttable}
\centering
\caption{Data sources for continuum bound-free (bf) and free-free (ff) absorption as well as scattering processes included in the \blue{} opacity code.
\label{tb:cont-source}}
{\begin{tabular*}{\columnwidth}{@{\extracolsep{\fill}}cc}
\toprule[2pt]
  Absorption/Scattering & Reference
  \\
\midrule[1pt]
  \ce{H-} bf  & \citet{McLaughlin_2017}
  \\ 
  \ce{H-} ff  & \citet{Bell_1987}
  \\
  \ce{H} bf & \citet{paul_barklem_2016_50215}
  \\
  \ce{H} ff & \citet{1988ApJ...327..477H}
  \\
  \ce{He-} ff & \citet{John_1995}
  \\
  \ce{Li-} ff & \citet{1975MNRAS.172..305J}
  \\
  \ce{C-} ff & \citet{Bell_1988}
  \\
  \ce{N-} ff & \citet{Ramsbottom_1992}
  \\
  \ce{O-}, \ce{Ne-}, \ce{Na-}, \ce{Ar-} ff & \citet{1975MNRAS.172..305J}
  \\
  \ce{Kr-}, \ce{Xe-}, \ce{Cs-}, \ce{Hg-} ff & \citet{1975MNRAS.172..305J}
  \\
  \ce{H2-} ff & \citet{Bell_1980}
  \\
  \ce{H2+} ff & \citet{1994ApJ...430..360S}
  \\
  \ce{N2-}, \ce{O2-}, \ce{CO-}, \ce{H2O-} ff & \citet{1975MNRAS.172..305J}
  \\
  \ion{He}{i}, \ion{He}{ii} bf & TOPbase (a)
  \\
  \ion{C}{i}, \ion{C}{ii}, \ion{C}{iii} bf & TIPbase (b)
  \\
  \ion{N}{i}, \ion{N}{ii}, \ion{N}{iii} bf & TIPbase
  \\
  \ion{O}{i}, \ion{O}{ii}, \ion{O}{iii} bf & TIPbase
  \\
  \ion{Ne}{i}, \ion{Ne}{ii}, \ion{Ne}{iii} bf & TOPbase
  \\
  \ion{Na}{i}, \ion{Na}{ii}, \ion{Na}{iii} bf & TOPbase
  \\
  \ion{Mg}{i}, \ion{Mg}{ii}, \ion{Mg}{iii} bf & TOPbase
  \\
  \ion{Al}{i}, \ion{Al}{ii}, \ion{Al}{iii} bf & TOPbase
  \\
  \ion{Si}{i}, \ion{Si}{ii} bf & TIPbase
  \\
  \ion{Si}{iii} bf & TOPbase
  \\
  \ion{S}{i} bf & TOPbase
  \\
  \ion{S}{ii}, \ion{S}{iii} bf & TIPbase
  \\
  \ion{Ar}{i}, \ion{Ar}{ii}, \ion{Ar}{iii} bf & TOPbase
  \\
  \ion{Ca}{i}, \ion{Ca}{ii}, \ion{Ca}{iii} bf & TOPbase
  \\
  \ion{Fe}{i}, \ion{Fe}{ii}, \ion{Fe}{iii} bf & TIPbase
  \\
  \ion{Ni}{ii} bf & TIPbase
  \\
\midrule[1pt]
  Electron scattering & \citet{thomson1912xlii}
  \\
  \ion{H}{i} Rayleigh scattering (c) & \citet{2004MNRAS.347..802L}
  \\
  \ion{He}{i} Rayleigh scattering (d) & \citet{PhysRevA.10.829}
  \\
\bottomrule[2pt]
\end{tabular*}}

    \begin{tablenotes}
      \item Note a: TOPbase is the Opacity Project online atomic database \citep{1992RMxAA..23..107C,1993A&A...275L...5C}, available at \url{https://cds.unistra.fr/topbase/topbase.html}
      \item Note b: TIPbase contains atomic data computed for the IRON Project \citep{1993A&A...279..298H}, available at \url{https://cdsweb.u-strasbg.fr/tipbase/home.html}
    \item Note c: For wavelengths $\lambda > 150$ nm. 
      \item Note d: For wavelengths $\lambda > 92$ nm.
    \end{tablenotes}
    
\end{threeparttable}
\end{table}

\begin{table}
\centering
\caption{Molecular species and corresponding line list data adopted in the \blue{} code. 
With the exception of \ce{H2O}, these data were drawn from the Kurucz database.
\label{tb:mol-source}}
{\begin{tabular*}{\columnwidth}{@{\extracolsep{\fill}}cc}
\toprule[2pt]
  Molecule & Reference
  \\
\midrule[1pt]
  \ce{AlH} & \citet{2018MNRAS.479.1401Y}
  \\ 
  \ce{AlO} & \citet{2015MNRAS.449.3613P}
  \\
  \ce{CaH} & \citet{2012MNRAS.425...34Y}
  \\
  \ce{CaO} & \citet{2016MNRAS.456.4524Y}
  \\
  \ce{C2} & \citet{2013JQSRT.124...11B}
  \\
  \ce{CN} & \citet{2014ApJS..210...23B}
  \\
  \ce{CO} & \citet{1992RMxAA..23...45K}
  \\
  \ce{CH} & \citet{2014Masseron}
  \\ 
  \ce{CrH} & \citet{2001Bauschlicher,2002ApJ...577..986B}
  \\
  \ce{FeH} & \citet{2003ApJ...594..651D}
  \\
  \ce{H2} & \citet{1992RMxAA..23...45K}
  \\
  \ce{MgH} & \citet{1992RMxAA..23...45K}
  \\
  \ce{MgO} & \citet{DAILY2002111}
  \\
  \ce{NaH} & \citet{2015MNRAS.451..634R}
  \\
  \ce{NH} & \citet{1992RMxAA..23...45K}
  \\
  \ce{OH} & \citet{1998JQSRT..59..453G,BERNATH200920}
  \\
    \ce{OH^{+}} & \citet{2018ApJ...855...21H}
  \\
  \ce{SiH} & \citet{2018MNRAS.473.5324Y}
  \\
  \ce{SiO} & \citet{1992RMxAA..23...45K}
  \\
  \ce{TiH} & \citet{2005ApJ...624..988B}
  \\
  \ce{TiO} & \citet{2019MNRAS.488.2836M}
  \\
  \ce{VO} & \citet{2016MNRAS.463..771M}
  \\
  \ce{H2O} & \citet{2006MNRAS.368.1087B}
  \\
\bottomrule[2pt]
\end{tabular*}}
\end{table}

We used the \blue{} code developed by \citet{2016MNRAS.463.1518A,2018A&A...615A.139A} to calculate total
line and continuous monochromatic extinction coefficients at different gas temperatures and densities.
The first step to calculating these quantities is to obtain the number density of each species from the EOS.
The original EOS implemented in \blue{} is based on the Saha ionisation equation 
albeit with a truncation of ionisation energies to account for non-ideal gas effects.
Although valid for studying the relatively cool 
photospheres of FGK-type stars,
a free-energy minimisation approach is expected to be more valid
in hotter environments \citep{1988ApJ...331..794H}.
Thus, in the high temperature regime\footnote{In 
this paper, temperatures at the stellar photosphere and lower atmosphere are called low temperatures. Quantitatively, we define low temperature as 
$\log(T / {\rm [K]}) < 4$, while $\log T > 4$ is referred as high temperature.}, we instead fed the number densities from \freeeos{} into \blue{}. For 
$\rm H^-$ whose number density is not accessible from \freeeos{}, we solved the Saha equation (assuming chemical equilibrium between neutral 
hydrogen and $\rm H^-$) to obtain its number density. 
It is worth mentioning that \freeeos{} handles only 20 elements whereas 83 elements are included in \blue{}. Therefore, for 63 elements that are 
not incorporated in the \freeeos{}, their number densities (including all ionisation states) were set to zero in \blue{} in this high temperature 
regime. Molecular abundances were also set to zero in this limit.

In the low temperature regime, we instead used the \blue{} EOS, so as to include
the effects of atoms and ions of all 83 elements, as well as of molecules.
The code uses molecular partition functions from \citet{2016A&A...588A..96B}.  Atomic and ionic partition functions
were calculated using data from the Kurucz online database \citep{1995ASPC...78..205K}, 
including at least three times ionised species for all the \freeeos{} metals except Cl (up to twice ionised), which is sufficient for the 
temperature range concerned in this work (i.e.~$\log(T / {\rm [K]}) \leq 4.5$). Our tests show that below $\log T = 4.5$, the number densities of fourth and 
higher ionised species are negligible for all 18 \freeeos{} metals.
In this work, for a given temperature, pressure, and chemical composition, the ionisation energy was truncated prior to the calculation of the partition function 
(by an amount calculated via the expression in Chapter 10.5 of \citealt{1999spa..book.....T}).

Given the EOS, \blue{} calculates total line and continuous monochromatic extinction coefficients using
transition cross-sections from various sources.
The bound-free and free-free data sources can be found in Table \ref{tb:cont-source}. We note that for many of species listed in Table \ref{tb:cont-source}, cross-sections of bound-free transitions were adopted from the Opacity project (TOP) or IRON Project (TIP) online database. These data are exactly the same as those used in  the previous generation of \stagger{} surface convection simulations \citep{2013A&A...557A..26M}.
We took data for bound-bound transitions in atoms, ions, and molecules from the Kurucz online database; we summarise the considered molecular species in Table \ref{tb:mol-source}.
With the exception of the scattering processes at the end of Table \ref{tb:cont-source}, all radiative transitions were
treated in true absorption.  
The line opacities are slightly influenced by the choice of microturbulence;
here we set this to $2\,\mathrm{km\,s^{-1}}$ as was used in the original \stagger{} grid.
In this work, for bound-bound transitions of species other than hydrogen, occupation probabilities $w$ were calculated following Appendix A of \citet{1994A&A...282..151H}.
Individual bound-bound transitions with monochromatic extinction coefficients $\alpha_{\nu;\mathfrak{m,n}}$ were
thus modified as $\alpha_{\nu;\mathfrak{m,n}}\rightarrow\alpha_{\nu;\mathfrak{m,n}}\times w_{\mathfrak{n}}$, where $\mathfrak{m}$ and $\mathfrak{n}$ denote the lower and
upper levels of the transition. 
For hydrogen lines and continua, we instead implemented the \texttt{HBOP} module provided by \citet{paul_barklem_2016_50215}.

\blue{} gives continuum and line absorption monochromatic extinction coefficients, which we added to get total 
absorption coefficients, separated into true absorption ($\alpha_{\rm ab}$) and scattering ($\alpha_{\rm sc}$).
These can also be added together to obtain a total extinction coefficient ($\alpha_{\rm tot}$).
We used these quantities in the opacity binning procedure described in Sect.~\ref{sec:op-bin}.

In order to have a smooth transition between the low and high temperature regimes, in practice we carried out two separate opacity calculations.
Extinction coefficients were computed using the \blue{} EOS in the low ($3.2 \leq \log T \leq 4.1$) regime, and
\freeeos{} in the high ($3.9 \leq \log T \leq 4.5$) temperature regime. 
The overlapping temperature interval $\log T \in [3.9,4.1]$ serves as an intermediate bridging region.
Here, we merged extinction coefficients from the two different calculations via
\begin{equation}
\begin{aligned}
f_{\rm b} &= \frac{1}{2} \left[ 
\sin \left( \frac{\log T - \log T_1}{\log T_2 - \log T_1} - \frac{1}{2} \right) \pi
+ 1 \right],
\\
\alpha_{\rm mg} &= \alpha_{\rm lowT} (1 - f_{\rm b}) + \alpha_{\rm highT} f_{\rm b},
\end{aligned}
\end{equation}
where $\log T_1$ and $\log T_2$ are the lower and upper boundary of the bridging region, respectively. The bridging function $f_{\rm b}$ equals 0 at the lower boundary, and smoothly increases to 1 at the upper boundary. The term $\alpha_{\rm mg}$ denotes the merged extinction coefficient, which could be monochromatic, or a mean value such as the Rosseland mean.

\subsubsection{Opacity grids} \label{sec:op-calc}

  Using the methods described in Sect.~\ref{sec:op-blue}, we constructed grids of monochromatic extinction coefficients for different values of $(\log T, \log\rho)$ for the AGSS09 and AAG21 solar chemical compositions. The temperature ranged from $\log T = 3.2$ to $4.5$ in steps of 0.01 $\log \mathrm{[K]}$, and the density ranged from $\log\rho = -13.7$ to $-1.0$ in steps of 0.1 $\log \mathrm{[g/cm^3]}$. This temperature-density coverage is sufficient for our applications since the surface convection simulations do not reach the low-density upper atmosphere nor extend to the high-temperature stellar interior.
  It is known that the carbon-to-oxygen ratio has a great impact on the molecular opacity when $\log T \lesssim 3.4$ \citep{2022ApJ...940..129M}. Nevertheless, given that the C/O ratio is very close between AGSS09 and AAG21, its influence on the opacity is limited as demonstrated in Fig.~6 of \citet{2022ApJ...940..129M}. The different Mg and Fe abundance in the two versions of solar composition, however, is likely to leave imprints on opacities as they are important electron donors
  (due to their high abundances and relatively low ionisation energies), and influence the \ce{H-} opacity (\citealt{2008A&A...486..951G} Sect.~6.3).  Although Si is another important electron donor, its abundance is identical in AGSS09 and AAG21.
  
  Extinction coefficients were computed at $250\,000$ wavelength points between 50 nm and 50 $\rm \mu m$, evenly sampled in logarithmic space to better resolve the ultraviolet. The resolving power is thus given by $\lambda / \Delta\lambda \approx 36\,200$. 
  This wavelength resolution is not adequate to resolve all absorption features caused by spectral lines. Nevertheless, in the scenario of stellar atmosphere modelling, the focus is to find a sufficient wavelength resolution such that the modelled temperature stratification converges rather than resolving all the line features in opacity calculation (see \citealt{2008PhST..133a4003P} Sect.~4 for detailed discussion). In order to verify our selected wavelength resolution, we have computed monochromatic extinction coefficients at very high wavelength resolution along $(\rho,T)$ points of the horizontal- and time-averaged \texttt{sunA09} model (cf.~Table.~\ref{tb:simu-info} and Sect.~\ref{sec:solar-model} about the model). The wavelength sampling is uniform in logarithm space, with two million points between 50 nm and 50 $\rm \mu m$, corresponding to a wavelength resolution $\lambda / \Delta \lambda \approx 289\,530$.
  We compared the temperature stratification predicted from high-resolution extinction coefficients with that adopted in this work using the 1D stellar atmosphere code \texttt{ATMO}. The \texttt{ATMO} code, described in \citet{2013A&A...557A..26M} Appendix A, employs the same EOS and opacity table as the \stagger{} code. At solar effective temperature and surface gravity, we found that 1D temperature structure evaluated from opacities with wavelength resolution $\lambda / \Delta \lambda \approx 289\,530$ and $\lambda / \Delta \lambda \approx 36\,200$ differ by less than 0.1\% in the stellar atmosphere, which translates to less than 5 K error in the optically thin regime. The error estimation implies that given the opacity binning method used in this work (cf.~Sect.~\ref{sec:op-bin}), our adopted wavelength sampling is sufficient for obtaining a reliable temperature structure.

We compared the Rosseland and Planck mean extinction coefficients with corresponding results from other opacity datasets, and found reasonable agreement in general (cf.~Appendix \ref{sec:compare-op} for quantitative comparisons).
The Rosseland mean extinction coefficient was calculated as
\begin{align} \label{eq:alphaRoss}
\alpha_{\rm Ross} = \frac{\int_{\lambda_{\min}}^{\lambda_{\max}} 
\frac{\partial B(T,\lambda)}{\partial T} \: d\lambda}
{\int_{\lambda_{\min}}^{\lambda_{\max}} 
\frac{1}{\alpha_{\rm tot}(\rho,T,\lambda)} 
\frac{\partial B(T,\lambda)}{\partial T} \: d\lambda}, 
\end{align}
and the Planck mean extinction coefficient as
\begin{align} \label{eq:alphaPl}
\alpha_{\rm Pl} = \frac{\int_{\lambda_{\min}}^{\lambda_{\max}} 
\alpha_{\rm tot} (\rho,T,\lambda) B(T,\lambda) \; d\lambda}
{\int_{\lambda_{\min}}^{\lambda_{\max}} B(T,\lambda) \; d\lambda},
\end{align}
where $B$ is the Planck function, 
\begin{equation}
B = \frac{2hc^2}{\lambda^5} \frac{1}{\exp\left(\frac{hc}{k_B T\lambda}\right) - 1},
\end{equation}
where $c$, $h$ and $k_B$ are the speed of light, Planck constant and Boltzmann constant, respectively. Numerically, the integrals in Eqs.~\eqref{eq:alphaRoss} and \eqref{eq:alphaPl} are discretised by the trapezoid rule, with the lower (upper) limit being the shortest (longest) wavelength point computed by \blue{}, which is $\lambda_{\min} = 50 \;\rm nm$ ($\lambda_{\max} = 50 \;\rm \mu m$) throughout our calculation. The thus evaluated mean extinction coefficients, together with continuum extinction coefficients (continuum absorption plus scattering) at 500 nm, are then interpolated to $(\rho,T)$ combinations that correspond to the EOS $(\rho,e_m)$ grid and stored in the EOS table as auxiliary quantities for the post-processing of simulation data (not used by the \stagger{} code).

\subsubsection{Opacity binning} \label{sec:op-bin}

\begin{figure}
\begin{overpic}[width=\columnwidth]{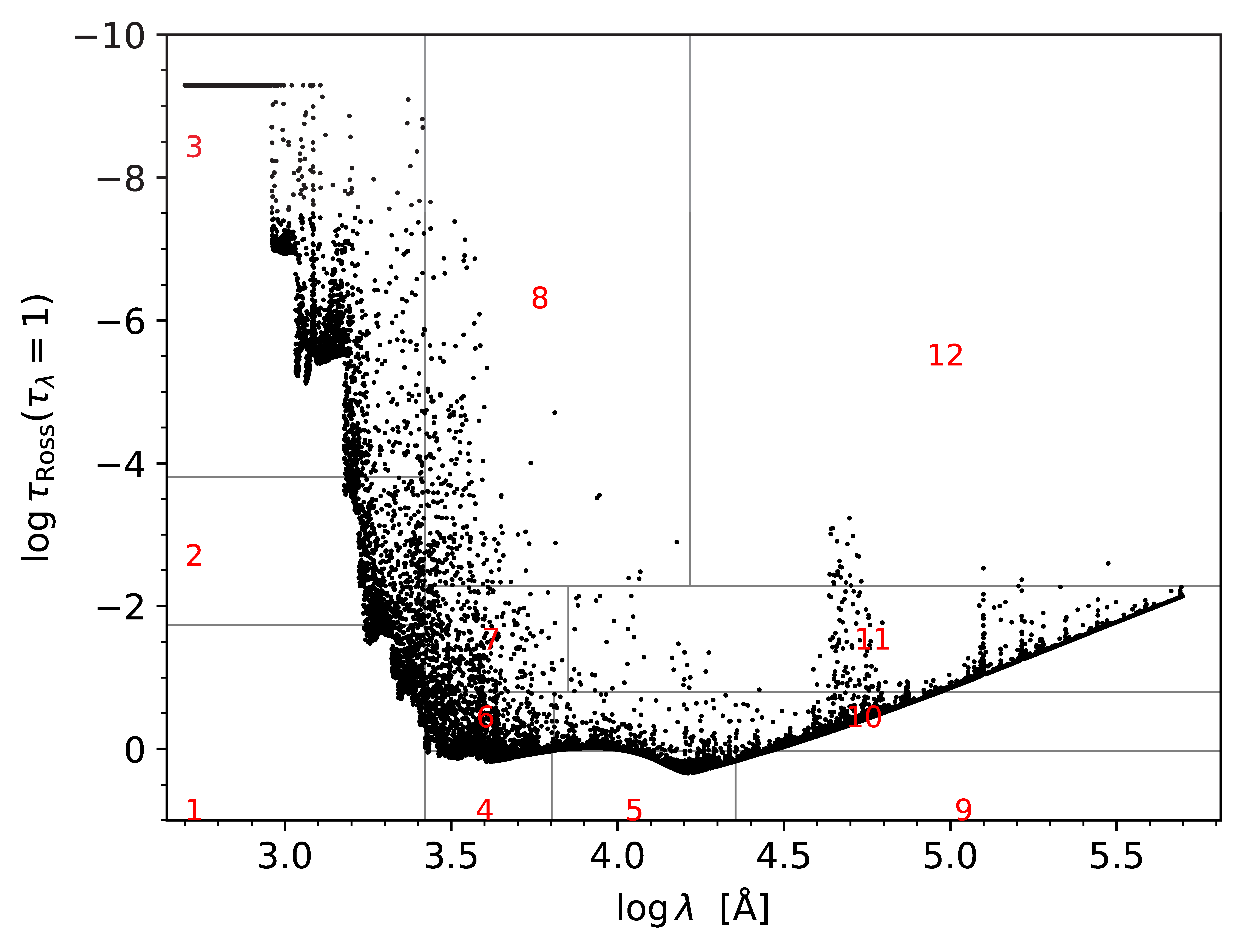}
\end{overpic}
\caption{The Rosseland optical depth where monochromatic optical depth is unity, which reflects the strength of opacity, as a function of wavelength computed based on the \mean{\rm 3D} \texttt{sunA09} model (cf.~Table \ref{tb:simu-info} and Sect.~\ref{sec:solar-model} for basic information about the model and how it is constructed) and \blue{} opacities assuming the AGSS09 solar abundance. One out of ten points in our wavelength sampling is shown to avoid over-crowded figure. Grey boxes numbered from 1 to 12 in red define the opacity bins used for the \texttt{sunA09} simulation. Bins 1-3 divide the ultraviolet wavelength region according to the location where flux emerges; Bins 4, 5 and 9 mainly consist of wavelength points distributed near the continuum-forming layers; Bins 8 and 12 include only wavelengths with strong opacity, which typically correspond to lines.
}
\label{fig:bin}
\end{figure}

  In radiative hydrodynamical simulations, solving the radiative transfer equation across the 3D simulation domain, at every time step, in about 10 different directions for a large number of wavelengths is computationally demanding. In order to make the problem computationally feasible, the \stagger{} code adopts the opacity binning method (also called the multi-group method, \citealt{1982A&A...107....1N,2000ApJ...536..465S,2013A&A...557A..26M,2018MNRAS.475.3369C}). 
  With this method we divide monochromatic opacities into multiple groups based on wavelength and opacity, or more precisely the approximate formation depth in a given model atmosphere. In each group, monochromatic opacities are appropriately averaged and treated as a ``single wavelength'' in the radiative transfer calculation, thereby reducing the workload enormously. We elaborate on our opacity binning procedure below and refer the readers to \citet[their Sect.~2.4.2]{2018MNRAS.475.3369C} for more information. A detailed study on different opacity binning approaches and their accuracy can be found in \citet{2023arXiv230603744P}.

\begin{figure*}
\begin{overpic}[width=0.99\textwidth]{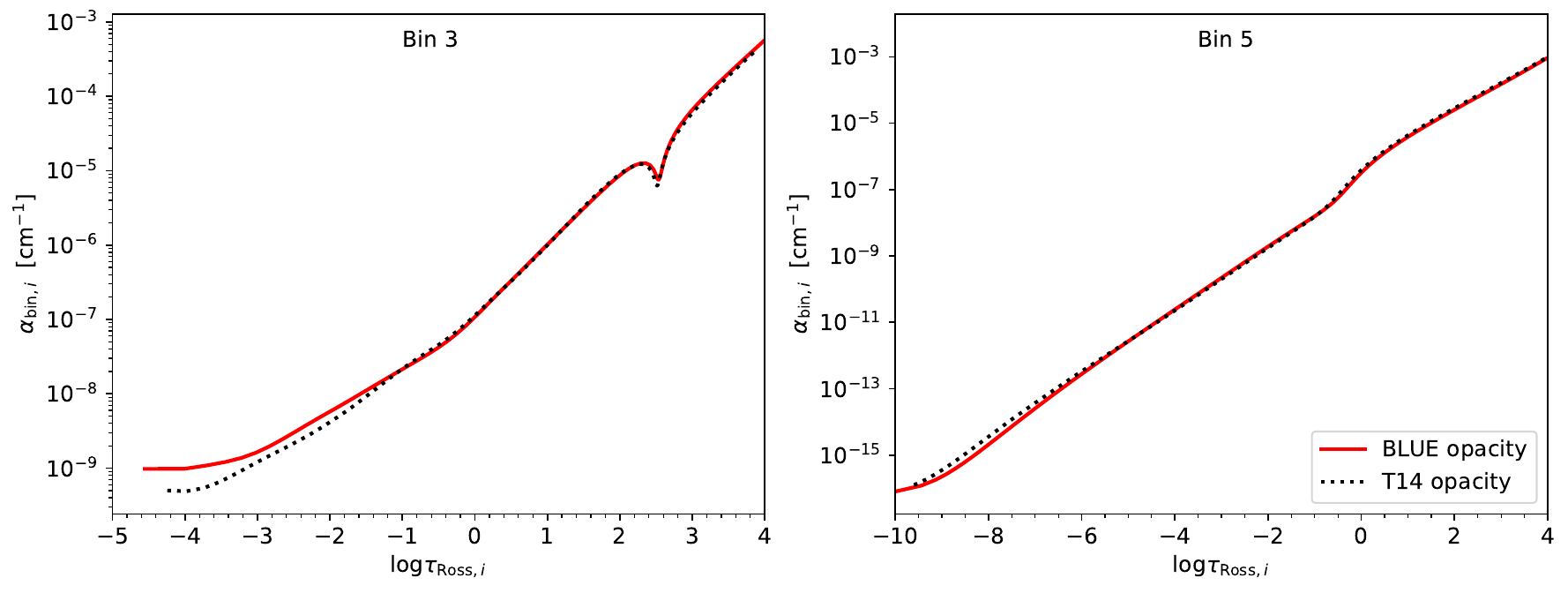}
\end{overpic}
\caption{Bin-averaged extinction coefficients (Eq.~\eqref{eq:alpha-bin}) as function of the bin-wise Rosseland optical depth of bins 3 and 5 (as defined in Fig.~\ref{fig:bin}), based on the \mean{\rm 3D} \texttt{sunA09} model. The bin-wise Rosseland optical depth $\tau_{{\rm Ross},i}$ is computed from wavelengths belonging to bin $i$.
Extinction coefficients calculated from the binning code implemented in \citet{2013ApJ...769...18T} with the \citet{2014MNRAS.442..805T} opacity dataset are shown in black dotted lines, whereas red solid lines represent $\alpha_{{\rm bin},i}$ used in this work. 
}
\label{fig:alpha_bin}
\end{figure*}

\begin{figure*}
\begin{overpic}[width=0.99\textwidth]{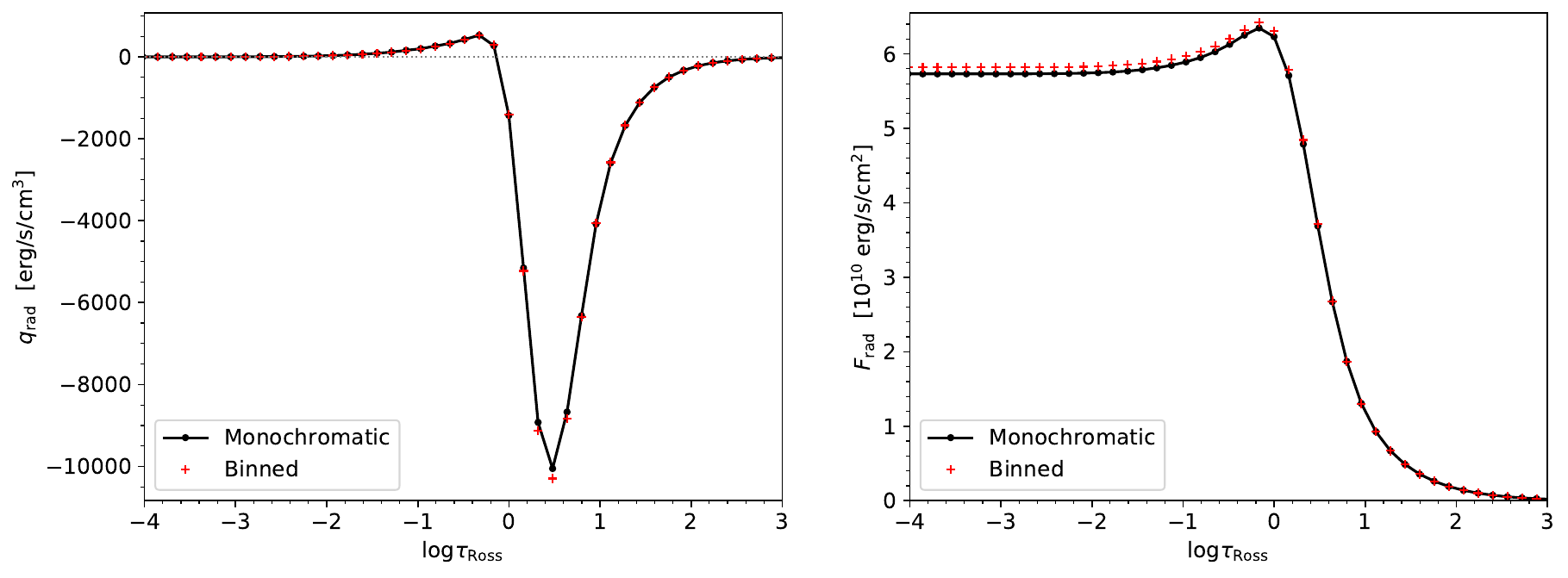}
\end{overpic}
\caption{Radiative heating (or cooling) rate and radiative flux computed based on the \mean{\rm 3D} \texttt{sunA09} model.
\textit{Left panel:} Comparison of the radiative heating (or cooling) rate computed from the 12 opacity bins shown in Fig.~\ref{fig:bin} (red plus mark) with the monochromatic solution using $250\,000$ wavelengths (black solid line) for the \mean{\rm 3D} \texttt{sunA09} model. The relative difference of $q_{\rm rad}$ at the cooling peak is 2.42\%. 
\textit{Right panel:} Comparison between radiative flux obtained from the opacity binning method and the monochromatic solution. The relative difference of surface flux (defined as the radiative flux at $\log\tau_{\rm Ross} = -4$) is 1.49\%.}
\label{fig:comp_mono_bin}
\end{figure*}

  Apart from the opacity data, a stellar atmosphere model is required for the binning process. In our implementation the horizontal- and time-averaged 3D model (\mean{\rm 3D} model hereinafter) was used, which implies opacity binning is an iterative process as the adopted \mean{\rm 3D} model affects the binned opacities, and the latter alters the stratification of the 3D atmosphere model in return. 
  Monochromatic absorption and total extinction coefficients as well as the Rosseland mean extinction coefficients computed at low and high temperature regions were first merged as described in Sect.~\ref{sec:op-calc} then interpolated to the densities and temperatures of the \mean{\rm 3D} model. Subsequently, we calculated the Rosseland optical depth $\tau_{\rm Ross}$ and monochromatic optical depth $\tau_{\lambda}$ according to the interpolated $\alpha_{\rm Ross}$ and $\alpha_{{\rm tot},\lambda}$, respectively. This is to obtain the Rosseland optical depth where monochromatic optical depth is unity, i.e.~$\tau_{\rm Ross}(\tau_{\lambda} = 1)$, which indicates the approximate location where flux emerges (also the approximate formation depth of lines). For a given selection of opacity bins demonstrated in Fig.~\ref{fig:bin}, all wavelength points were assigned to an opacity bin based on their wavelength and $\tau_{\rm Ross}(\tau_{\lambda} = 1)$ value.
  We note that the organisation (not the exact location of boundaries) of opacity bins in Fig.~\ref{fig:bin} is the same as what was adopted in the 3D solar model of \citet{2013A&A...554A.118P}, which was well-tested against various observational constraints.
  
  For each bin, averaged extinction coefficients were computed by integrating over wavelengths that belong to that bin. In the optically thick region, the Rosseland mean, $\alpha_{{\rm Ross},i}$, as defined in Eq.~\eqref{eq:alphaRoss} but integrating over the wavelengths and transitions 
  included in the $i$th bin,
  is a good representation of the mean extinction coefficient within that bin.
  
  In the optically thin part, however, the radiative flux cannot be described by the diffusion equation. Inspired by the fact that the divergence of monochromatic radiative flux is proportional to $\alpha_{{\rm tot},\lambda} (J_{\lambda} - B_{\lambda})$ in LTE, where $J$ denotes the monochromatic mean intensity (Eq.~\eqref{eq:Jlmd}), and considering that the absorption processes (characterized by $J$) are usually stronger than emission processes (characterized by the source function $B$ in LTE) in stellar atmospheres (see e.g.~Fig.~5 of \citealt{2012MNRAS.427...27B}), we used $J$ as the weighting function in the optically thin part \citep{1990A&A...228..155N,2013A&A...557A..26M}. 
  Also, scattering was excluded from the extinction coefficient when calculating mean opacities in the optically thin region. This is referred to as the \textit{no-scattering-in-streaming-regime} approximation. The purpose of this modification is to approximate the temperature structure predicted by surface convection simulations with continuum scattering processes properly treated in the radiative transfer calculation using simulations with LTE radiative transfer \citep{2011A&A...528A..32C}. Although the inclusion of continuum scattering in the extinction coefficient has little impact on the temperature structure of the 3D solar model (\citealt{2010A&A...517A..49H} Fig.~7), \citet{2011A&A...528A..32C} demonstrated that for metal-poor stars, the \textit{no-scattering-in-streaming-regime} approximation gives good agreements with the correct solution where  scattering is included in the modelling, whereas using the total extinction coefficient in the optically thin region overheats the stellar atmosphere. In order to be consistent with future non-solar metallicity simulations, the \textit{no-scattering-in-streaming-regime} approximation is adopted in this work. The same approximation was also used in the construction of the \stagger{}-grid (\citealt{2013A&A...557A..26M} Sect.~2.1.5).
The mean extinction coefficient in the optically thin regime is therefore
\begin{equation} \label{eq:alphaJ-bin}
\alpha_{J,i} = \frac{\int_{\lambda \in {\rm bin} \; i} \alpha_{\rm ab} J \: d\lambda}
{\int_{\lambda \in {\rm bin} \; i} J \: d\lambda},
\end{equation}
with $i = 1,2,...,12$ indicates a specific opacity bin. Note that, as discussed above, scattering is excluded from the integrand. The integration is carried out with the midpoint (or rectangle) method. We have verified that the midpoint integration rule is the preferred choice for the opacity binning problem, as it reproduces very well $\alpha_{J,i}$ and $\alpha_{{\rm Ross},i}$ obtained from monochromatic opacities with high wavelength resolution for all opacity bins. Higher-order integration methods such as the trapezoid and Simpson's rule, however, will lead to large errors in the mean extinction coefficient for some bins and result in incorrect temperature structure in the solar atmosphere.

  The two types of mean extinction coefficients were blended together with an exponential bridging function to obtain the final combined bin-averaged extinction coefficient, which reads
\begin{equation} \label{eq:alpha-bin}
\alpha_{{\rm bin}, i} = e^{-2\tau_{{\rm Ross},i}} \alpha_{J,i} + 
(1 - e^{-2\tau_{{\rm Ross},i}}) \alpha_{{\rm Ross},i},
\end{equation}
with $\alpha_{{\rm Ross},i}$ being the Rosseland mean extinction coefficient evaluated from wavelengths belonging to bin $i$ and $\tau_{{\rm Ross},i}$ the optical depth based on $\alpha_{{\rm Ross},i}$.

  The bin-averaged extinction coefficients for two selected opacity bins are depicted in Fig.~\ref{fig:alpha_bin}. 
  These are typical of opacity bins including optical and near-infrared wavelengths that form around the optical surface, and of opacity bins including ultraviolet wavelengths that form in higher layers.
  Red solid lines stand for $\alpha_{{\rm bin}, i}$ used in this work, whereas black dotted lines represent extinction coefficients calculated from the binning code implemented in \citet{2013ApJ...769...18T} with the \citet{2014MNRAS.442..805T} opacity dataset (see also Appendix \ref{sec:compare-op}), which corresponds to binned extinction coefficients adopted in previous \stagger{} simulations. 
  The $\alpha_{{\rm bin}, i}$ results from the previous and our new calculation agree reasonably well for both bin 3 and bin 5. The small difference seen in Fig.~\ref{fig:alpha_bin} is due to different opacity datasets adopted, as the opacity binning procedure is identical between this work and \citet{2013ApJ...769...18T}.

To get the mean intensity in Eq.~\eqref{eq:alphaJ-bin}, we solved the radiative transfer equation in the 1D plane parallel \mean{\rm 3D} model under the assumption of LTE:
\begin{equation} \label{eq:RT}
\mu\frac{dI_{\lambda}}{d\tau_{\lambda}} = I_{\lambda} - B_{\lambda},
\end{equation}
 where $\mu = \cos\theta$ represents the polar angle along which the equation is solved, $I_{\lambda}$ is the monochromatic intensity, and $\tau_{\lambda}$ is the vertical monochromatic optical depth.
Integrating $I_{\lambda}$ over the polar angle gives the mean intensity,
\begin{equation} \label{eq:Jlmd}
J_{\lambda} = \frac{1}{2} \int_{-1}^1 I_{\lambda} \; d\mu.
\end{equation}
The 1D LTE radiative transfer problem was solved using a modified \citet{1964CR....258.3189F} technique developed by \citet[their Sect.~3.9.1]{1982A&A...107....1N} and \citet[their Sect.~4]{2003ASPC..288..519S}, which rearranges the Feautrier transport equation and solves for $Q_{\lambda} = (1/2) [I_{\lambda}(\mu) + I_{\lambda}(-\mu)] - B_{\lambda}$ in order to improve the numerical accuracy at large optical depths. This is the same numerical technique as is implemented in the 3D radiative transfer solver of the \stagger{} code. The integration in Eq.~\eqref{eq:Jlmd} was approximated with the Gaussian-Legendre quadrature with five polar angles, which is sufficient for this problem as increasing the number of $\mu$-angles hardly changes the resulting mean intensity.

  Reducing a large number of wavelengths to 12 opacity bins in the radiative transfer calculation is a significant simplification. To examine how accurate the opacity binning method is, we compared the radiative heating (or cooling) rate computed from bin-averaged quantities
\begin{equation}
q_{\rm rad,b} = 4\pi \sum_{i=1}^{12} \alpha_{{\rm bin},i} (J_i - B_i)
\end{equation}
with the monochromatic solution
\begin{equation}
q_{\rm rad,m} = 4\pi \int_{\lambda_{\min}}^{\lambda_{\max}} 
\alpha_{{\rm tot}, \lambda}(J_{\lambda} - B_{\lambda}) \: d\lambda.
\end{equation}
Here $B_i = \int_{\lambda \in {\rm bin} \; i} B_{\lambda} \: d\lambda$ is the Planck function integrated over a given bin, and $J_i$, the mean intensity of bin $i$, was obtained from Eqs.~\eqref{eq:RT} and \eqref{eq:Jlmd} with source function $B_i$ and optical depth computed via $\alpha_{{\rm bin},i}$. The heating rate is a crucial outcome of the radiative transfer process because it enters directly into the energy equation thereby influencing the temperature stratification of the model. The difference in $q_{\rm rad}$ is quantified by $\max\vert q_{\rm rad,b} - q_{\rm rad,m} \vert / \max\vert q_{\rm rad,m} \vert$ \citep{2013A&A...557A..26M}, which is the relative difference at the cooling peak in most cases.
  In addition to $q_{\rm rad}$, we examined how well the opacity binning method reproduces the radiative flux, as it determines the effective temperature of the model. The radiative flux ($F_{\rm rad}$) was calculated by integrating the heating rate from the bottom (stellar interior) to the top (atmosphere) along the \mean{\rm 3D} model. Thus, one could realise that differences in $F_{\rm rad}$ are a manifestation of differences in $q_{\rm rad}$. Nevertheless, comparing $F_{\rm rad}$ probes the mean differences in $q_{\rm rad}$ rather than at a particular location. In short, we employed a combination of $\max\vert q_{\rm rad,b} - q_{\rm rad,m} \vert / \max\vert q_{\rm rad,m} \vert$ and the relative difference in $F_{\rm rad}$ as an indicator of the accuracy of opacity binning:
\begin{equation}
\Delta_{\rm b,m} = \frac{\max\vert q_{\rm rad,b} - q_{\rm rad,m} \vert}{\max\vert q_{\rm rad,m} \vert} + 
\left\vert \frac{F_{\rm rad,b} - F_{\rm rad,m}}{F_{\rm rad,m}} \right\vert. 
\end{equation}
As a criterion of the realism of the opacity binning method, $\Delta_{\rm b,m}$ was utilised to select the ``best'' binning configuration for a given model atmosphere and opacity data: The ``best selection'' of opacity bins corresponds to the global minimum of $\Delta_{\rm b,m}$. In practice, we iteratively adjusted the location of bin boundaries and computed corresponding $\Delta_{\rm b,m}$. The optimisation problem was tackled with Powell's method (cf., for example, \citealt{1992nrfa.book.....P} Sect.~10.5) with the location of bin boundaries as multi-dimensional variables and $\Delta_{\rm b,m}$ the minimisation target. 
  Our preferred bin selection for model \texttt{sunA09} (Table \ref{tb:simu-info}) obtained from minimising $\Delta_{\rm b,m}$ is illustrated in Fig.~\ref{fig:bin}. 
  Because \blue{} monochromatic extinction coefficients for the AGSS09 and AAG21 abundance are close to each other at most wavelengths, the optimised location of bin boundaries for model \texttt{sunA21} is nearly identical to that of the \texttt{sunA09} model.
  We caution that for multi-variable optimisation, it is very challenging to find the global minimum, therefore our preferred bin selections might represent only the local minimum of $\Delta_{\rm b,m}$ and affected by our initial guess.
  
  A comparison of $q_{\rm rad}$ and $F_{\rm rad}$ between the opacity binning and monochromatic calculation is presented in Fig.~\ref{fig:comp_mono_bin}. In the case of \mean{\rm 3D} \texttt{sunA09} model, the relative difference of $q_{\rm rad}$ at the cooling peak (in this case the same as $\max\vert q_{\rm rad,b} - q_{\rm rad,m} \vert / \max\vert q_{\rm rad,m} \vert$) is 2.42\%, which is of similar accuracy level as \citet{2013A&A...557A..26M}. The relative difference of surface flux (defined as the radiative flux at $\log\tau_{\rm Ross} = -4$) is 1.49\%, which translates to a 0.37\% relative difference and $\approx 21$ K absolute difference in effective temperature.
  Nevertheless, we note that errors in surface flux presented here are merely illustrative. Owing to the non-linear, turbulent nature of 3D surface convection simulations, it is possible that the true error of binning is larger than the estimation based on the $\mean{\rm 3D}$ model. True errors in flux can be determined by synthesising the flux spectrum with 3D model atmosphere and comparing it with observation.

\section{Solar atmosphere model} \label{sec:solar-model}

\begin{table*}
\centering
\begin{threeparttable}
\caption{Fundamental parameters and basic information about solar simulations presented in this study. 
All simulations listed below adopt the same surface gravity and mesh properties.
We note that the effective temperature fluctuates over time in 3D models, therefore both the mean effective temperature and its standard deviation are presented. Also, both minimum and maximum vertical grid spacing are provided, as mesh points are not uniformly distributed vertically. Nominal values of the solar effective temperature and surface gravity are $T_{\rm eff,\odot} = 5772$ K, $\log g_{\odot} = 4.438$ (cgs unit) \citep{2016AJ....152...41P}.
\label{tb:simu-info}}
{\begin{tabular*}{0.6\textwidth}{@{\extracolsep{\fill}}lccc}
\toprule[2pt]
  Model name            & \texttt{sunA09}   & \texttt{sunA21}   & \texttt{sunRef} 
  \\
\midrule[1pt]
  $T_{\rm eff}$ (K)     &  $5775 \pm 15$    &  $5778 \pm 15$    &  $5773 \pm 16$
  \\
  $\log g$ (cgs)        & 4.438             & 4.438             & 4.438
  \\
  Chemical composition  & AGSS09            & AAG21             & AGSS09
  \\
  EOS                   & \freeeos{}        & \freeeos{}        & MHD (a)
  \\
  Opacity               & \blue{}           & \blue{}           & MARCS (b)
  \\
\midrule[1pt]
  Numerical resolution          & \multicolumn{3}{c}{$240^2 \times 230$}
  \\
  Time duration (min)           & \multicolumn{3}{c}{200}
  \\
  Sampling interval (s)         & \multicolumn{3}{c}{30}
  \\
  Vertical size (Mm)            & \multicolumn{3}{c}{3.6}
  \\
  Vertical grid spacing (km)    & \multicolumn{3}{c}{7--33}
  \\
  Horizontal grid spacing (km)  & \multicolumn{3}{c}{25}
  \\
\bottomrule[2pt]
\end{tabular*}}

    \begin{tablenotes}
      \item Note a: Customised version as described in \citet{2013ApJ...769...18T}.
      \item Note b: Customised version as described in \citet{2010A&A...517A..49H}. 
    \end{tablenotes}
    
\end{threeparttable}
\end{table*}

  The EOS and opacities described in Sect.~\ref{sec:physics} were incorporated in the \stagger{} code as basic input physics for our 3D solar model atmospheres. The \stagger{} code \citep{1995...Staggercodepaper,2018MNRAS.475.3369C,stein:StaggerCode} is a radiation-magnetohydrodynamics code that solves the time-dependent equation of mass, momentum and energy conservation, the magnetic-field induction equation, as well as the radiative transfer equation on a 3D staggered Eulerian mesh. 
  Solar models in this study have been constructed without magnetic fields.
  Radiative energy transport was modelled in LTE. The equation of radiative transfer with the Planck function as source function (Eq.~\eqref{eq:RT}) was solved with a modified \citet{1964CR....258.3189F} technique \citep{1982A&A...107....1N,2003ASPC..288..519S} for all mesh points above $\tau_{\rm Ross} = 500$ at every time-step of the simulation. The frequency dependence of the radiative transfer problem was approximated by the opacity binning method detailed in Sect.~\ref{sec:op-bin}, where the layout of 12 opacity bins was optimised individually for each solar model. Spatially, the radiative transfer equation was solved along nine different directions which consist of one vertical and eight inclined directions representing the combination of two polar angles and four azimuthal angles. The integration over the polar angle was approximated by the Gauss-Radau quadrature. The thus evaluated radiative heating rate enters the equation of energy conservation and meanwhile was used to compute the radiative flux and the effective temperature of the model.
  
  The two new 3D solar atmosphere models presented in this work are labelled \texttt{sunA09} and \texttt{sunA21}. The former adopts the AGSS09 solar chemical composition while the latter uses the recent AAG21 abundance. Both model atmospheres were constructed based on the reference solar effective temperature and surface gravity given by \citet{2016AJ....152...41P}. Their basic configurations are summarised in Table \ref{tb:simu-info}. In addition, in the subsequent sections we present comparisons of these two models with that of an older \stagger{} model (i.e.~with the same input physics as used in the \stagger{}-grid).  This model has been used in previous studies (e.g.~\citealt{2019ApJ...880...13Z}). We refer to as \texttt{sunRef} hereafter.
  
  The simulation domain is discretised on a Cartesian mesh located around the solar photosphere (with coordinates $x,y,z$ where $y$ denotes the vertical dimension). For both models, the distribution of mesh is identical to that used in \texttt{sunRef}.  The horizontal extent of the simulations is 6 Mm $\times$ 6 Mm with 240 mesh points evenly distributed in each direction, which is large enough to enclose at least ten granules at any time of the simulation \citep{2013A&A...557A..26M}. 
  There are 240 mesh points in the vertical direction, where five layers at the top and bottom of the simulation domain are reserved as the so-called ``ghost-zone'' to ensure that vertical boundary conditions fulfil the six-order numerical differentiation scheme employed in the \stagger{} code (\citealt{1995...Staggercodepaper} Sect.~2.2). The remaining 230 vertical meshes constitute the ``physical domain'' of the simulation and will be equated with the simulation domain hereinafter.
  The vertical size of our simulations is 3.6 Mm excluding ghost zones, which extends from 2.7 Mm below the base of the photosphere (the near-surface convection layers) to 0.9 Mm above it (the bottom of the chromosphere). This corresponds roughly to the outer 0.5\% of the solar radius. Because the vertical scale of the simulations is tiny compared to the solar radius, spherical effects are negligible and the surface gravity can be used in the entire simulation domain. The 230 vertical mesh points are not evenly placed: the finest numerical resolution is applied around the optical surface in order to resolve the steep transition from the optically thick to thin regime (see \citealt{2013A&A...557A..26M} Fig.~2 for an illustration). Given the size of the simulation box, the $240^2 \times 230$ numerical resolution was verified to be adequate for line formation calculations \citep{2000A&A...359..669A} which is the main application of our models. 
  
  The boundaries are periodic in the horizontal directions, while open in the vertical \citep{2018MNRAS.475.3369C}. At the bottom boundary, outgoing flows (vertical velocities towards the stellar centre) freely carry their entropy fluctuations out of the simulation domain, whereas constant entropy and thermal (gas plus radiation) pressure is enforced for incoming flows. Temporally, our simulations span 200 solar minutes, with one snapshot stored every 30 seconds solar time. All these simulation snapshots were generated after numerical relaxation procedures described in Sect.~2.3 of \citet{2013A&A...557A..26M}.
  
  We note that except for the updates to the input physics, the setup and mesh properties of \texttt{sunA09} and \texttt{sunA21} simulations are practically identical to model \texttt{sunRef}, which is well-tested against other 3D solar atmosphere models and observational constraints \citep[hereafter P2013]{2012A&A...539A.121B,2013A&A...554A.118P}.\footnote{To be precise, the \stagger{} solar model atmosphere presented in \citet{2012A&A...539A.121B} and P2013 is slightly different from model \texttt{sunRef}: The former two were generated with an older version of the \stagger{} code and uses the \citet{2005ASPC..336...25A} chemical composition, while the latter adopts the AGSS09 abundance. Nevertheless, these two models are identical in all other aspects and the differences in their photospheric structure are minor. 
  \label{ftnt:OtherModels}}
  To this end, we verify the new atmosphere models by comparing them with model \texttt{sunRef} in Sects.~\ref{sec:mean-3D} and \ref{sec:distri}.

\subsection{Spatially and temporally averaged model} \label{sec:mean-3D}

\begin{figure*}
\includegraphics[width=0.49\textwidth]{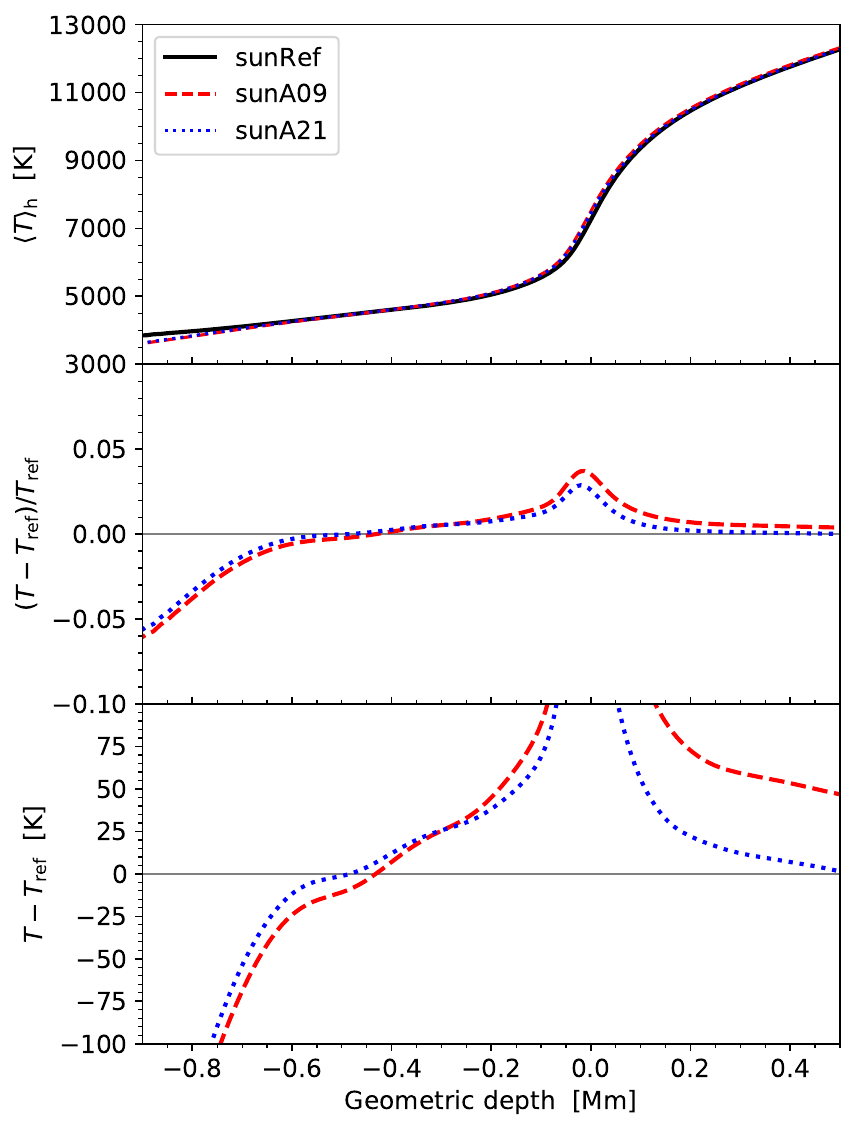}
\includegraphics[width=0.49\textwidth]{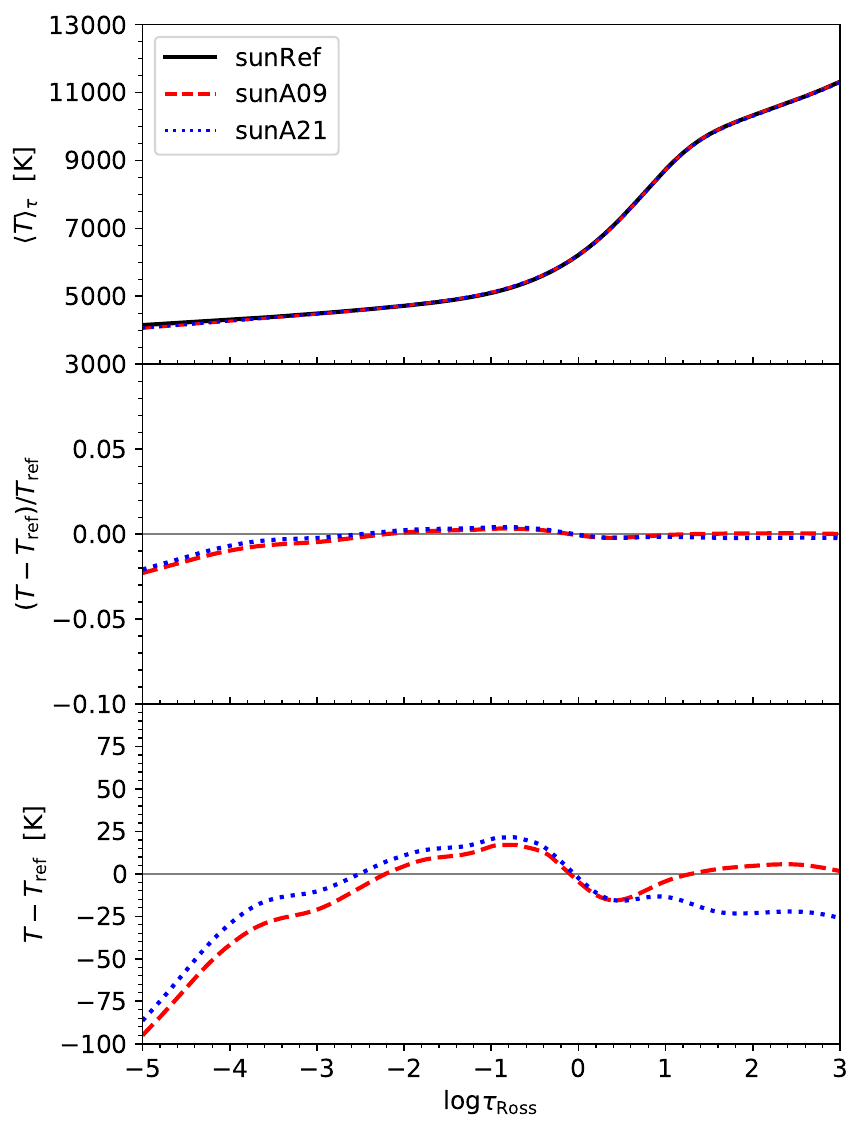}
\caption{Averaged temperature profiles for different solar model atmospheres.
\textit{Left panel:} Simple horizontal- and time-averaged temperature profiles $\mean{T}_{\rm h}$ as a function of geometric depth. Zero geometric depth corresponds approximately to the optical surface. Relative and absolute differences between the new models and \texttt{sunRef} are shown in the middle and lower part, respectively.
\textit{Right panel:} The $\tau_{\rm Ross}$- and time-averaged temperature $\mean{T}_{\tau}$ as a function of Rosseland optical depth ($T - \tau$ relation) for models \texttt{sunA09}, \texttt{sunA21} and \texttt{sunRef}.}
\label{fig:comp_T}
\end{figure*}
\begin{figure*}
\subfigure{
\begin{overpic}[width=0.49\textwidth]{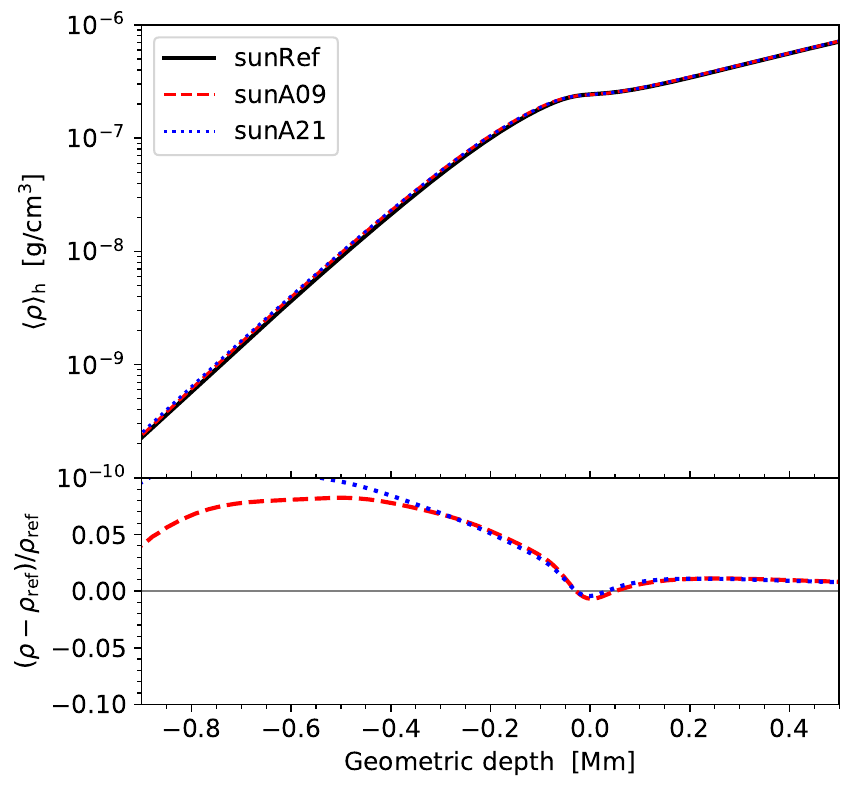}
\end{overpic}
}
\subfigure{
\begin{overpic}[width=0.49\textwidth]{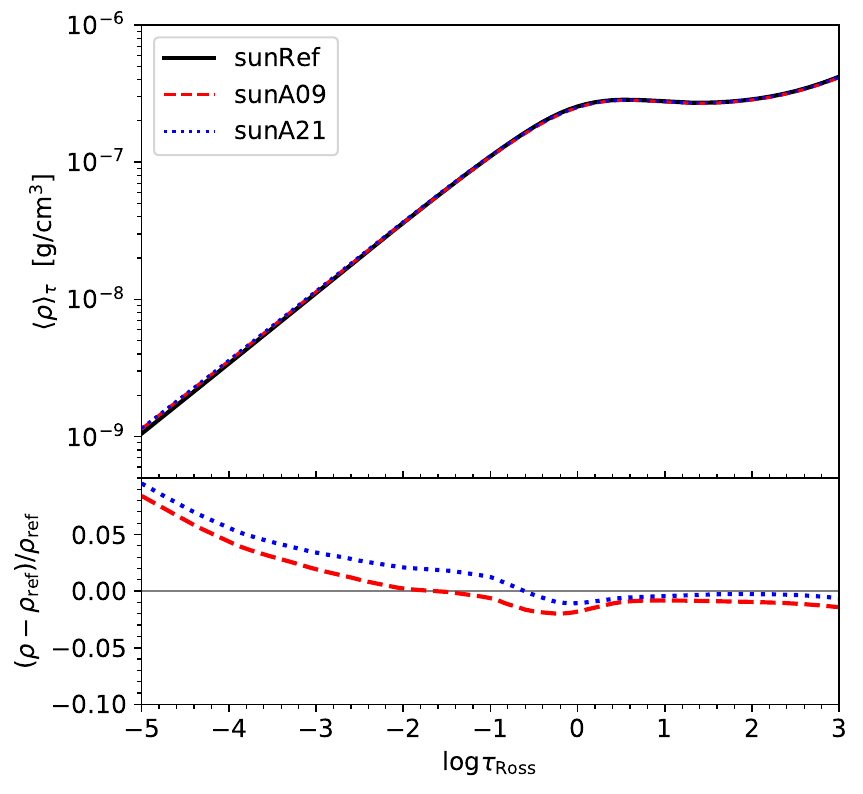}
\end{overpic}
}
\caption{Similar to the \textit{top} and \textit{middle panels of} Fig.~\ref{fig:comp_T}, but showing the mean density stratification for different solar simulations.
}
\label{fig:comp_rho}
\end{figure*}
\begin{figure*}
\subfigure{
\begin{overpic}[width=0.49\textwidth]{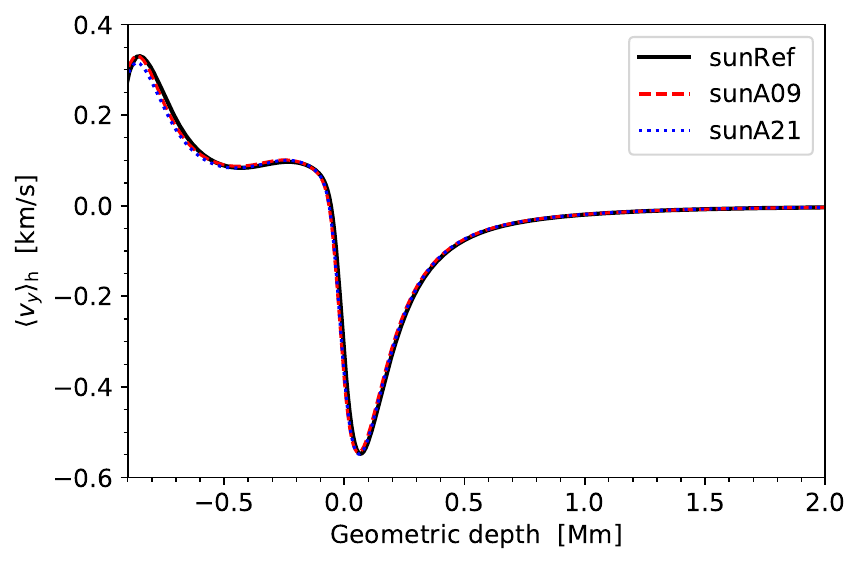}
\end{overpic}
}
\subfigure{
\begin{overpic}[width=0.49\textwidth]{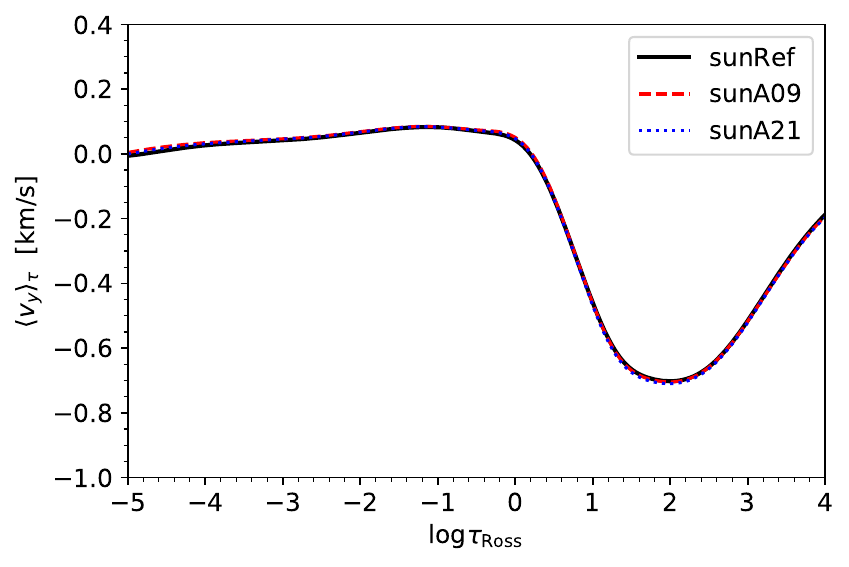}
\end{overpic}
}
\caption{The temporal mean of simple horizontal-averaged (\textit{left panel}) and $\tau_{\rm Ross}$-averaged vertical velocity (\textit{right panel}) for different solar simulations. Positive velocities correspond to downflow that move towards the stellar interior.}
\label{fig:comp_vy}
\end{figure*}
\begin{figure*}
\subfigure{
\begin{overpic}[width=0.49\textwidth]{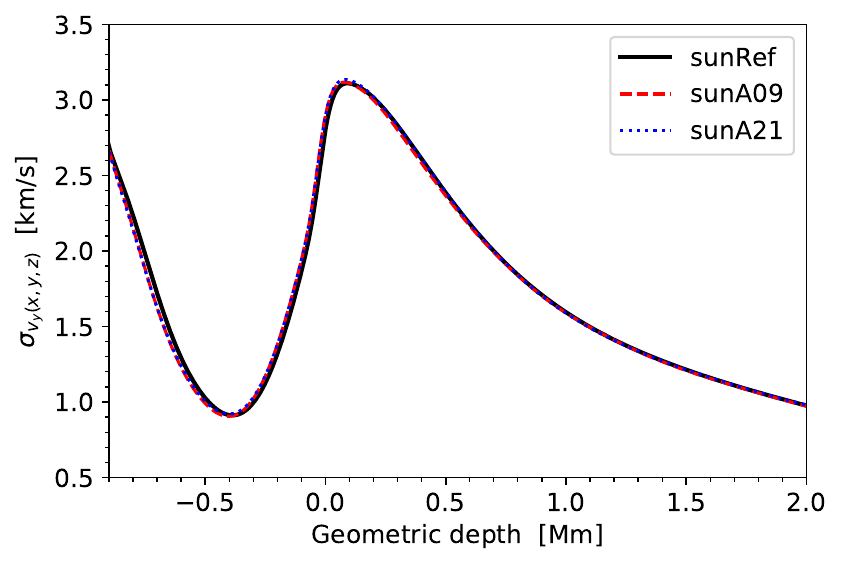}
\end{overpic}
}
\subfigure{
\begin{overpic}[width=0.49\textwidth]{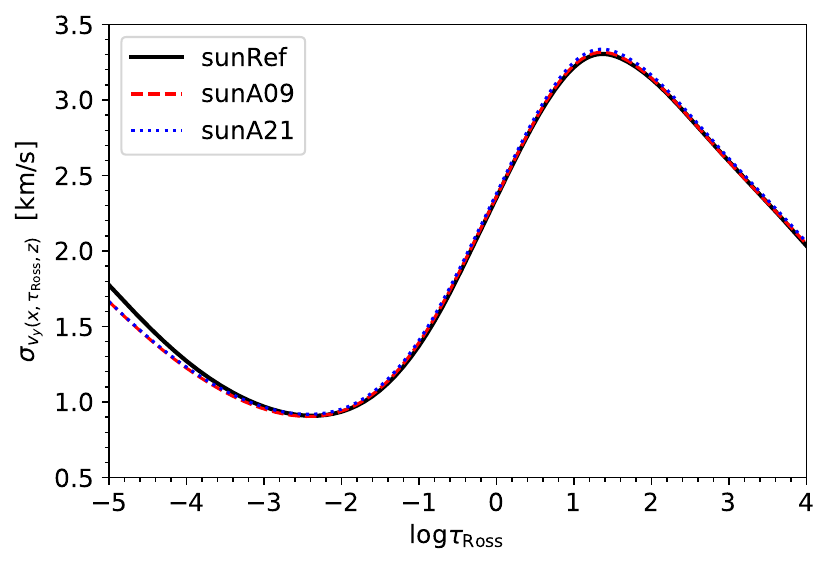}
\end{overpic}
}
\caption{The time-averaged standard deviation of vertical velocity in geometric depth scale (\textit{left panel}) and Rosseland optical depth scale (\textit{right panel}). This quantity, also called the root-mean-square (rms) of vertical velocity fluctuation, indicates the variation of vertical velocity at a given depth.}
\label{fig:comp_vyrms}
\end{figure*}
\begin{figure}
\begin{overpic}[width=\columnwidth]{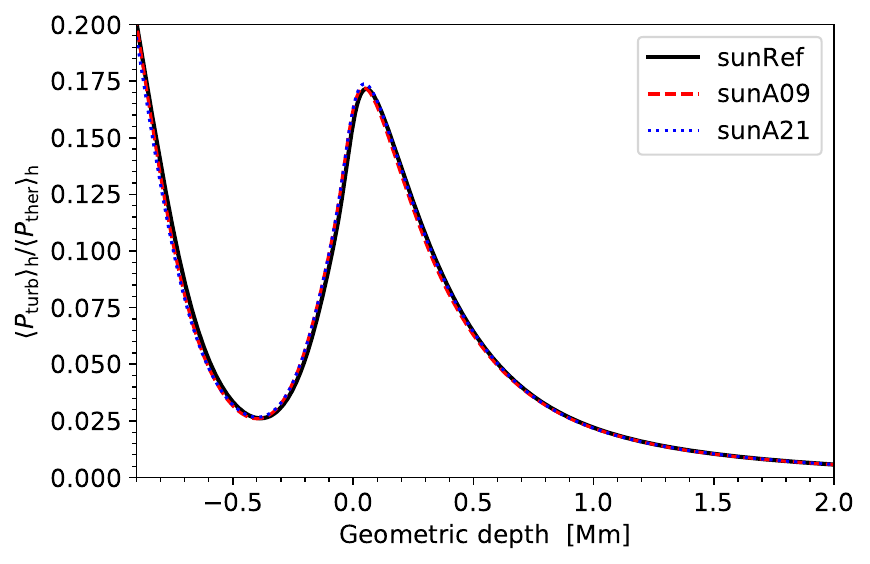}
\end{overpic}
\caption{Ratio of simple horizontal- and time-averaged turbulent pressure to thermal pressure.}
\label{fig:comp_Pratio}
\end{figure}
\begin{figure}
\begin{overpic}[width=\columnwidth]{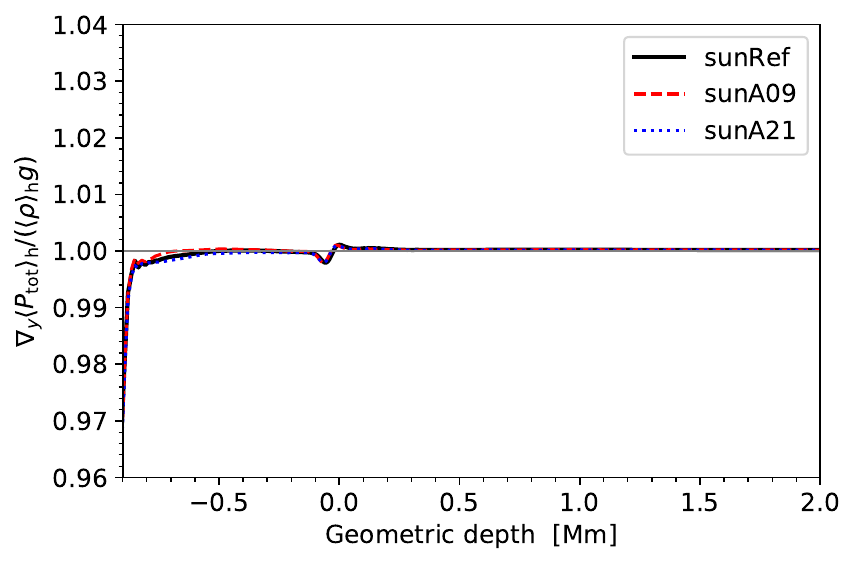}
\end{overpic}
\caption{Deviations from hydrostatic equilibrium as a function of vertical geometric depth for $\rm \mean{3D}_h$ models.}
\label{fig:comp_HE}
\end{figure}

  Fig.~\ref{fig:comp_T} shows the mean temperature structure for the two new models as well as model \texttt{sunRef}. Directly comparable are the \texttt{sunA09} and \texttt{sunRef}, as they are based on the same solar abundance. 
  Here, two different methods were used when averaging over space -- the simple horizontal average and the average over layers of constant Rosseland optical depth ($\tau_{\rm Ross}$-average). The simple horizontal averaged quantities were obtained by taking the mean value at given vertical geometric depths, in practice defined by the numerical mesh. The $\tau_{\rm Ross}$-average was achieved by first computing the Rosseland optical depth for the entire simulation domain. For an arbitrary physical quantity $f$, this establishes an $f - \tau_{\rm Ross}$ relationship at every column of the simulation box. For all columns, the $f(\tau_{\rm Ross})$ function was then interpolated to a reference optical depth frame. Taking the mean value for all interpolated $f$ at a particular reference optical depth gives the $\tau_{\rm Ross}$-averaged quantity. Spatially averaged quantities were then averaged over the whole time series of the simulation, i.e., every simulation snapshot, to obtain the horizontal- and time-averaged model. In this paper, we use symbols $\mean{...}_{\rm h}$ and $\mean{...}_{\tau}$ to represent the spatial and temporal averaging over constant vertical geometric depth and Rosseland optical depth, respectively. 
  
  We note that there are other ways to average 3D models. However, the focus of this section is to compare the mean structure of new models with the reference model \texttt{sunRef}: We aim neither to compare different averaging methods nor to determine the suitable averaging method for a certain application. We refer the readers to \citet{2013A&A...560A...8M} for a thorough investigation in this direction. 

  We can see from the \textit{middle} and \textit{bottom left panel} of Fig.~\ref{fig:comp_T} that around zero geometric depth, $\mean{T}_{\rm h}$ of \texttt{sunA09} and \texttt{sunRef} differ by more than 100 K (relative difference about 3\%). This discrepancy arises because model \texttt{sunA09} and \texttt{sunRef} (also \texttt{sunA21}) adopt identical geometric depth scale but are computed with different opacities. Therefore, their optical surfaces correspond to slightly different geometric depths. Due to the large temperature gradient in the convective-radiative transition zone, a small mismatch in the placement of the optical surface will cause a considerable temperature difference in the geometric depth scale. To this end, a more sensible approach is to compare the averaged temperature profile based on the optical depth scale.
  From the \textit{right panels} of Fig.~\ref{fig:comp_T}, it is clear that \texttt{sunA09} and \texttt{sunRef} have similar mean temperature structure in general: their $\tau_{\rm Ross}$- and time-averaged temperature differs by less than 25 K above $\log\tau_{\rm Ross} \sim -3$. The absolute temperature differences reach up to $\sim 100$ K at $\log\tau_{\rm Ross} = -5$. Nevertheless, the highly turbulent outermost layers of the simulation are likely the least realistic given our neglect of magnetic fields.

  In Appendix \ref{sec:compare-T}, we isolate the impact of EOSs on the mean temperature structure of 3D solar models by constructing two nearly identical models that differ solely in their input EOS. It turns out that using MHD or \freeeos{} will lead to $\sim 15$ K temperature difference in optically thick layers. However, in most parts of the optically thin regime, averaged temperatures between models with MHD and \freeeos{} agree within 5 K, suggesting that temperature differences shown in the \textit{right panels} of Fig.~\ref{fig:comp_T} are primarily attributed to different opacity data and the selection of opacity bins between \texttt{sunA09} and \texttt{sunRef}. 
  Here we emphasise that a careful selection of opacity bins is of great importance to obtain a reliable temperature structure. From our experience, different binning configurations could  affect the averaged temperature profile by more than 50 K in the modelled solar atmosphere. Although the exact number depends on the location of the atmosphere where the temperature difference is measured, the impact of binning configuration is clearly non-negligible, to a degree that is much stronger than the EOS effect especially in the optically thin region. 
  We note that \citet{2018MNRAS.475.3369C} modelled the atmosphere of a metal-poor red giant with the opacity binning method and reached the same conclusion that an erroneous selection of bins leads to about 100 K temperature discrepancy in the stellar atmosphere (see their Fig.~9).

  As shown in Fig.~\ref{fig:comp_vy}, the three solar models give similar mean vertical velocity profiles. Particularly notable is the large upward, mean vertical velocity just below the photosphere, which is a consequence of surface convection. The upflows and downflows that form the observed solar granulation pattern must have the same absolute momentum. 
  We can construct a toy model that assumes all upflows (downflows) have identical density $\rho_{\rm up}$ ($\rho_{\rm dn}$) and vertical velocity $v_{\rm up}$ ($v_{\rm dn}$), leading to an equation for the conservation of momentum, 
\begin{equation} \label{eq:momem-cons-toy}
a \rho_{\rm up} v_{\rm up} + b \rho_{\rm dn} v_{\rm dn} = 0.
\end{equation}
Here $a$ and $b$ is the fractional area covered by upflows and downflows (the ``filling factor'') respectively. A simple rearrangement of Eq.~\eqref{eq:momem-cons-toy} gives
\begin{equation} \label{eq:momem-cons-alt}
a v_{\rm up} + b v_{\rm dn} = 
-b \left( \frac{\rho_{\rm dn}}{\rho_{\rm up}} - 1 \right) v_{\rm dn}.
\end{equation}
The left hand side of Eq.~\eqref{eq:momem-cons-alt} is the mean vertical velocity. Because the density is typically higher in downflows (\citealt{1998ApJ...499..914S} Fig.~10), the mean vertical velocity has the same direction as the upflows as depicted in Fig.~\ref{fig:comp_vy}. The magnitude of $\mean{v_y}_{\rm h}$ reflects the asymmetry between upflows and downflows. The strongest asymmetry is found in the convective region just below the optical surface, where the vertical velocity fluctuations, represented by $\sigma_{v_y}$, are also the largest (Fig.~\ref{fig:comp_vyrms}).

The ratio of turbulent to thermal pressure, which is a proxy for vertical velocity fluctuations, is demonstrated in Fig.~\ref{fig:comp_Pratio}. Thermal pressures were evaluated from the EOS while turbulent pressures were computed via (\citealt{1999A&A...351..689R} Sect.~3)
\begin{equation}
P_{\rm turb} = \rho \left( v_y - \frac{\overline{\rho v_y}}{\overline{\rho}} \right)^2,
\end{equation}
where overlines stand for horizontal (but not temporal) averaging. At most vertical layers, the three solar models agree well in terms of $\mean{P_{\rm turb}}_{\rm h} / \mean{P_{\rm ther}}_{\rm h}$. 

  For our plane-parallel radiative-hydrodynamical simulations, horizontally averaged fluid properties averaged over sufficiently long stellar time should fulfil the equation of hydrostatic equilibrium (\citealt{2013A&A...560A...8M} Appendix A.2). We check how close the $\rm \mean{3D}_h$ models are to hydrostatic equilibrium in Fig.~\ref{fig:comp_HE}, and it turns out that hydrostatic equilibrium is fulfilled at most parts for all three solar simulations. However, we observe deviations from hydrostatic equilibrium in the uppermost layers for all solar models considered, which indicates momentum is not conserved at the top boundary. Nevertheless, the top boundary has little impact on the stratification of the 3D model because of the low density there.
  Meanwhile, we note that it is the total (thermal plus turbulent) pressure that enters into the equation of hydrostatic equilibrium, as detailed in \citet{2013A&A...560A...8M}.

\subsection{Distribution of intensity and vertical velocity} \label{sec:distri}

\begin{figure}
\begin{overpic}[width=\columnwidth]{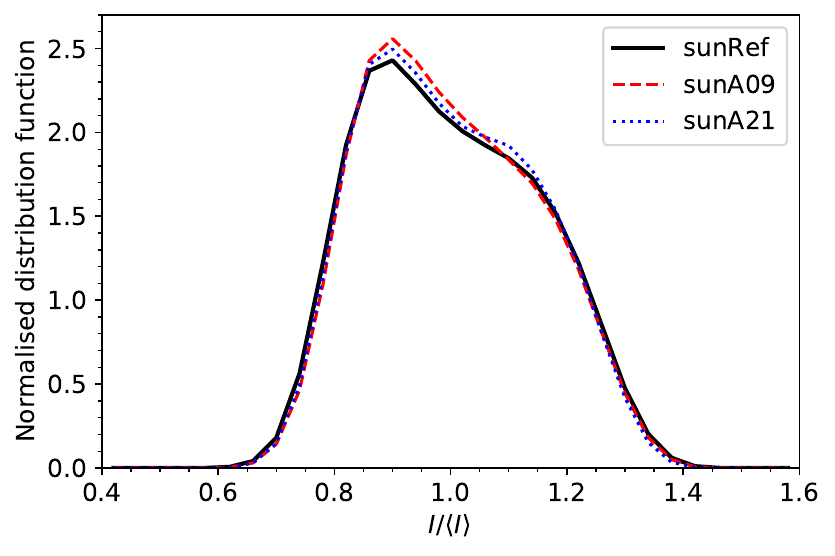}
\end{overpic}
\caption{Time-averaged distribution of disk-centre bolometric intensity predicted by the two solar models with new input physics but different chemical compositions, \texttt{sunA09} (AGSS09 composition) and \texttt{sunA21} (AAG21 composition), as well as the reference
model with unmodified input physics, \texttt{sunRef} (AGSS09 composition). The bolometric intensity is normalised by its mean value $\mean{I}$, and the distribution function is normalised such that its integration over $I / \mean{I}$ equals one.}
\label{fig:comp_histI}
\end{figure}

\begin{figure*}
\subfigure{
\begin{overpic}[width=0.53\textwidth]{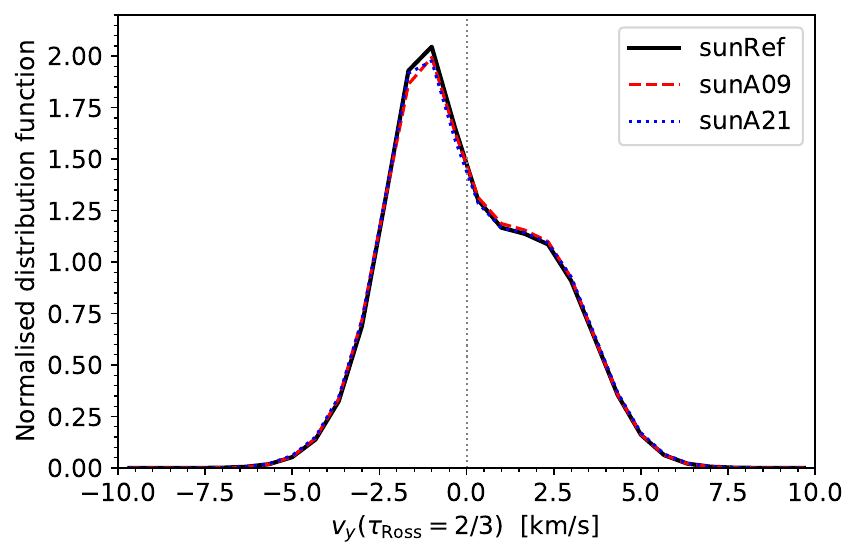}
\end{overpic}
}
\subfigure{
\begin{overpic}[width=0.45\textwidth]{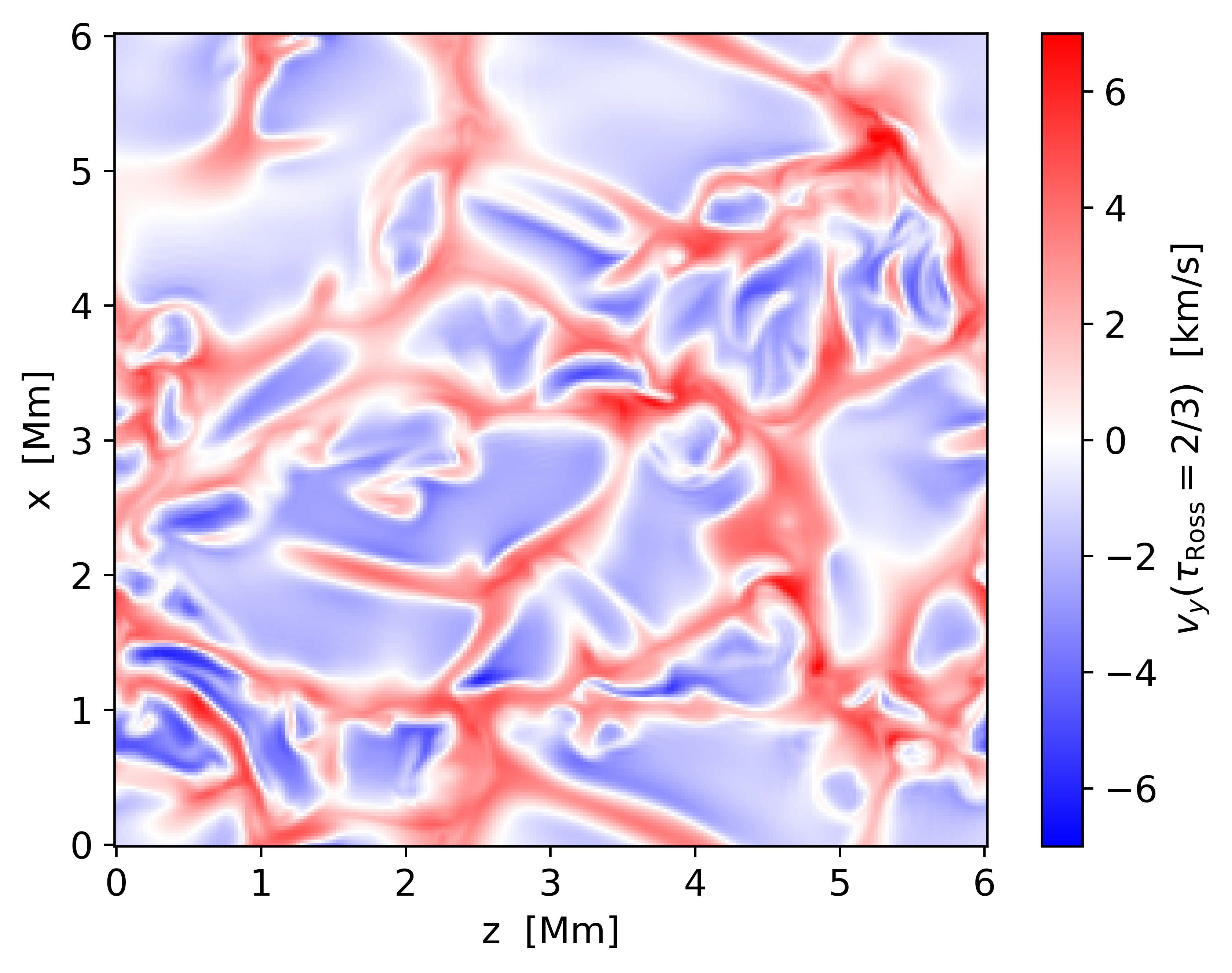}
\end{overpic}
}
\caption{Characteristics of the surface vertical velocity predicted by 3D solar models.
\textit{Left panel:} Time-averaged distribution of surface vertical velocity. The histograms were calculated based on 30 equidistant bins between $-10$ and 10 km/s, where positive velocities indicate downflow moving towards the stellar interior. The distribution function was normalised in the same way as the bolometric intensity (Fig.~\ref{fig:comp_histI}) so that the total area under it equals one.
\textit{Right panel:} Spatially resolved surface vertical velocity pattern from one snapshot of the \texttt{sunA09} simulation.}
\label{fig:comp_histvy}
\end{figure*}

  Checking the mean stratification provides an intuitive overview of 3D models, but meanwhile wipes out fluctuations across the horizontal plane. In this section, we will scrutinize the distribution of key simulation properties at selected horizontal planes, which captures the inhomogeneity in the convective motions.

  One of the main breakthroughs brought by surface convection simulations is that they revealed how convection operates in the convective-radiative boundary layers of stars. In the photosphere, fluid elements rapidly lose their heat to radiation and become denser than their surroundings. The overdense material is pulled down by negative buoyancy through the optical surface forming the intergranular lanes. Below the surface, conservation of mass forces the lower-density, warmer material to rise back through the optical surface, forming the so-called granules (cf.~\citealt{2009LRSP....6....2N} for detailed description). 
  The distribution of emergent intensity is a direct reflection of the radiation field in granules and intergranular lanes, which originated from upflows and downflows at different heights of the atmosphere.
  
  Here we compare the disk-centre bolometric intensity distribution of \texttt{sunA09} and \texttt{sunA21} with the reference model \texttt{sunRef}. The distribution of bolometric intensity across the simulation domain is shown as a histogram of normalised intensity $I/\mean{I}$, where $\mean{I}$ is the mean bolometric intensity. In all cases, 30 equidistant bins were assigned between $I/\mean{I} = 0.4$ and 1.6 for the evaluation of the distribution function. The time-averaged distribution shown in Fig.~\ref{fig:comp_histI} was obtained by computing the normalised distribution function for every simulation snapshot then averaging over all snapshots. The intensity distribution of the new solar models agrees well with model \texttt{sunRef}, all showing a bimodal distribution with a primary peak located at $I/\mean{I} \approx 0.9$ that corresponds to intergranular lanes, and a secondary peak at a higher intensity $I/\mean{I} \approx 1.1$. However, the new models predict slightly higher peaks around $I/\mean{I} = 0.9$.
  
  The area coverage and the strength of upflows and downflows is revealed by the distribution of vertical velocities. For each simulation snapshot, vertical velocities at each column were interpolated to $\tau_{\rm Ross} = 2/3$ to obtain the velocity distribution at the vicinity of the optical surface.\footnote{For the Eddington grey atmosphere, the location of the optical surface is strictly $\tau = 2/3$. However, this is generally not exact for realistic model atmospheres.} Averaging over all distribution functions gives the time-averaged velocity distribution shown in Fig.~\ref{fig:comp_histvy}. Similar to the case of intensity, the distribution function of vertical velocity appears to be bimodal, where the primary peak corresponds to upflow. It is worth noting that the distribution function confirms the visual impression of the right panel of Fig.~\ref{fig:comp_histvy} that upflow fills more area in the simulation domain.

\section{Comparison with observations} \label{sec:compare-obs}
 
  The best way of examining the fidelity of stellar models is to compare model predictions with observables. In this section, we compute the absolute flux spectrum (Sect.~\ref{sec:flux-spec}), the centre-to-limb variations (Sect.~\ref{sec:CLV}) and hydrogen lines (Sect.~\ref{sec:Hlines}) from the new solar model atmospheres. All modelling results are compared with solar observations as well as theoretical predictions presented in P2013, which is based on a well-established solar atmosphere model computed with the \stagger{} code and the \citet{2005ASPC..336...25A} abundance. We name this model \texttt{sunP2013} to avoid confusion with model \texttt{sunRef} mentioned in Sect.~\ref{sec:solar-model} (see footnote \ref{ftnt:OtherModels}).

  The spectrum synthesis was carried out using the 3D non-LTE radiative transfer code \balder{} \citep{2018A&A...615A.139A}, a branch of \multitd{} \citep{2009ASPC..415...87L} with updates for example to the formal solver \citep{2016MNRAS.463.1518A,2019A&A...624A.111A} and in particular to the EOS and opacities as discussed in Sect.~\ref{sec:op-blue}.  In this work, identical abundances and opacity data were employed in \balder{} and the surface convection simulation.
  The calculations follow what was presented in \citet{2018A&A...615A.139A} and employ the same model atom.
  Previous investigations have indicated that departures from LTE have non-negligible effects on the wings of Balmer lines (particularly \ce{H\alpha}, see e.g., Fig.~7 of P2013 and Fig.~4 of \citealt{2018A&A...615A.139A}): In the solar case, Balmer lines computed in non-LTE show weaker wings than in LTE.

\subsection{Absolute flux spectrum} \label{sec:flux-spec}

\begin{figure}
\begin{overpic}[width=\columnwidth]{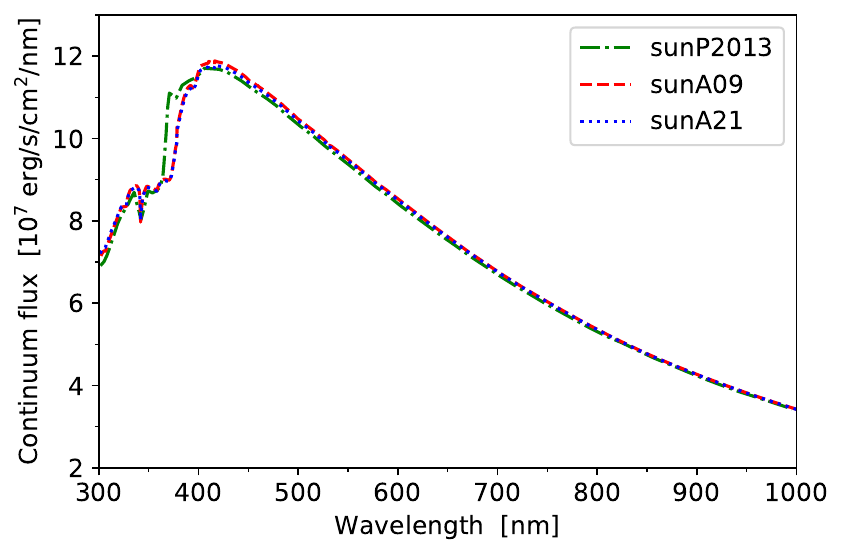}
\end{overpic}
\caption{Continuum emergent flux spectra computed from new 3D solar atmosphere models (red dashed line and blue dotted line indicate results from \texttt{sunA09} and \texttt{sunA21}, respectively). The theoretical result from model \texttt{sunP2013} is also shown in the green dash-dotted line. All fluxes presented here are absolute values at the solar surface.
}
\label{fig:comp_flux_cont}
\end{figure}

\begin{figure*}
\includegraphics[width=0.99\textwidth]{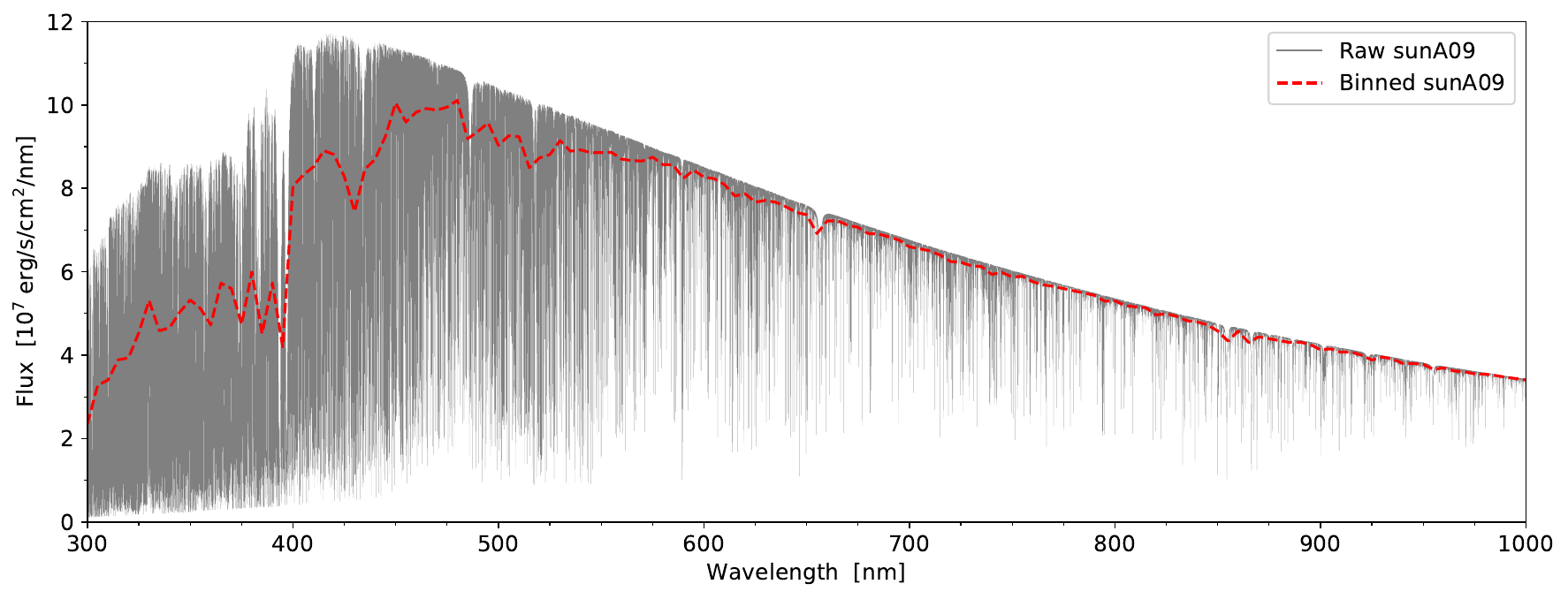}
\includegraphics[width=0.49\textwidth]{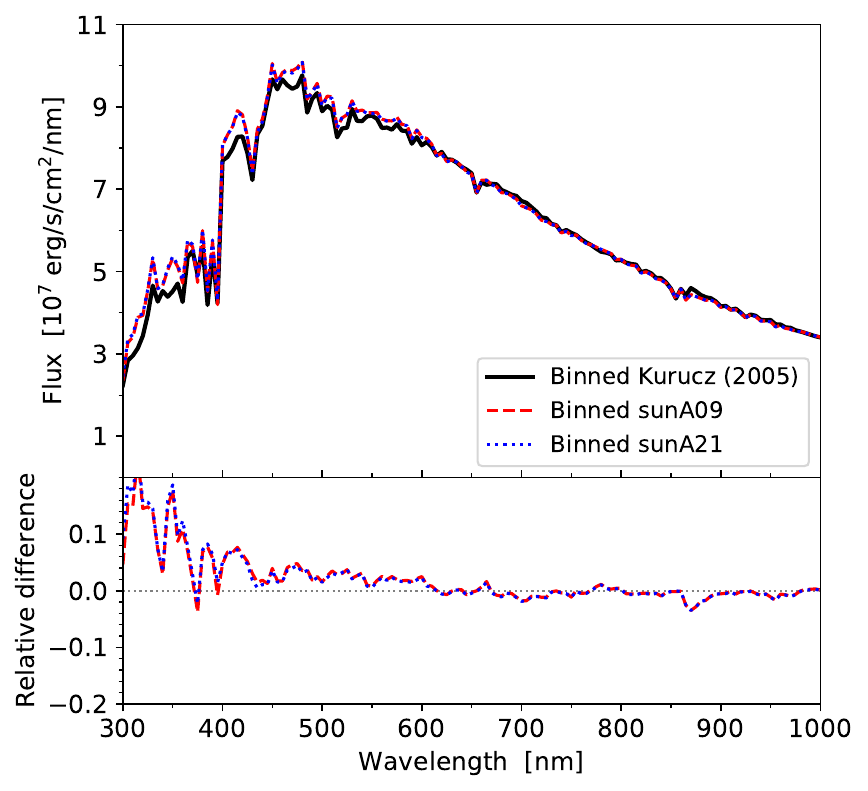}
\includegraphics[width=0.49\textwidth]{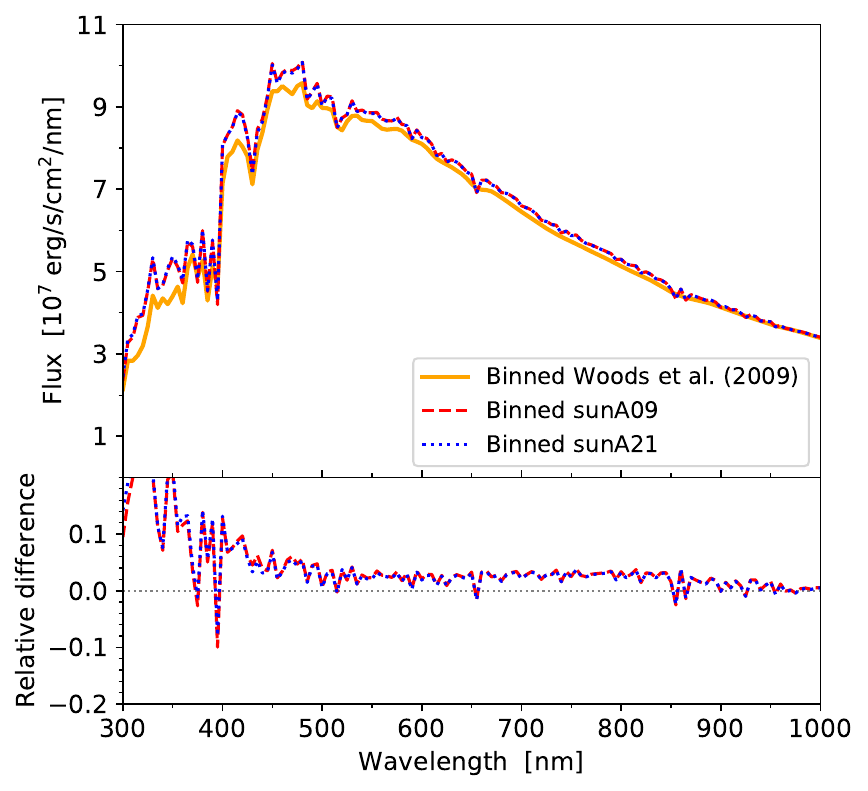}
\caption{The distribution of the solar absolute emergent flux from 300 to 1000 nm.
\textit{Upper panel:} The thin grey line is the absolute emergent flux computed using \balder{} based on the \texttt{sunA09} model, at a resolution of $\lambda / \Delta\lambda = 50000$. The smoothed synthetic spectrum using a 5 nm  wavelength bin is depicted in red dashed line (equivalent to the red dashed line in the \textit{right panel}).
\textit{Lower left panel:} Comparison of smoothed synthetic spectra with the smoothed solar flux of \citet[black solid line]{2005MSAIS...8..189K}. All results presented here were smoothed with a 5 nm wavelength bin in order to facilitate comparison between simulation and observation. Relative differences in smoothed absolute flux as a function of wavelength, evaluated through $F_{\lambda} / F_{\lambda,\rm obs} - 1$ with $F_{\lambda,\rm obs}$ denoting the smoothed measured values, are shown in the \textit{bottom right panel} for two solar models. 
\textit{Lower right panel:} Similar to the \textit{lower left panel}, but the smoothed synthetic spectra are compared with the smoothed Solar Irradiance Reference Spectra of \citet[orange solid line]{2009GeoRL..36.1101W}.
All fluxes presented here are absolute values at the solar surface.
}
\label{fig:full_spec}
\end{figure*}

  The emergent flux spectrum (or spectral energy distribution) plays an important role in stellar physics. Theoretical flux spectra generated from model atmospheres can be applied to, for example: calculate synthetic photometry \citep{2014MNRAS.444..392C,2018A&A...611A..11C}, determine stellar parameters \citep{2022MNRAS.513.2719V} and derive interstellar extinctions \citep{2023ApJS..264...41Y}. Previous investigations have demonstrated that 3D model atmospheres are able to produce realistic absolute flux spectra for the Sun \citep{2018A&A...611A..11C,2018A&A...613A..24K}. It is therefore worth checking how our new models perform in this respect. 
  
  Here we first compare the continuum flux spectrum predicted by our new models with that of model \texttt{sunP2013}. Fig.~\ref{fig:comp_flux_cont} shows that except for wavelengths slightly above the Balmer jump ($\approx 364.5$ nm), continuum flux spectra computed from the two new models and \texttt{sunP2013} agree well with each other, indicating the temperature stratification of the three models are close to each other around the optical surface.
  The differences to the red of the Balmer jump can be attributed to the treatment of dissolved Rydberg states as implemented in the \texttt{HBOP} module of \citet{paul_barklem_2016_50215}, that lead to a smooth decay of the continuous opacity instead of a sharp transition.
  This is also apparent in the continuum centre-to-limb variation discussed in Sect.~\ref{sec:CLV}.

  Comparing the synthesised continuum flux with observation is challenging owing to the difficulty of deriving the continuum level from irradiance data in the ultraviolet wavelength region (cf.~\citealt{1984SoPh...90..205N} Sect.~5 and P2013 Sect.~4). To this end, we elect to synthesise the absolute flux spectrum by incorporating the information of spectral lines into opacities used in the radiative transfer calculation and comparing with the solar irradiance data of \citet{2005MSAIS...8..189K} as well as the Solar Irradiance Reference Spectra of \citet{2009GeoRL..36.1101W}. The latter was measured during the solar minimum in 2008. 
  For both \texttt{sunA09} and \texttt{sunA21} models, we computed the theoretical absolute flux spectrum with \balder{} using identical opacity data employed in our 3D atmosphere modelling. The flux spectrum calculation was carried out from 300 to 1000 nm, at a wavelength resolution of $\lambda / \Delta\lambda = 50\,000$. The thus obtained absolute flux spectrum is illustrated in the \textit{upper panel} of Fig.~\ref{fig:full_spec} for model \texttt{sunA09}.
  Nevertheless, the absolute flux spectrum contains a forest of lines, impeding detailed comparison between simulation and observation. Therefore, we heavily smooth both the synthetic and the observed spectra using a 5 nm wavelength bin such that line features in the spectra are smoothed out. Comparing the smoothed absolute spectra examines the temperature structure of the 3D model, which sets the modelled continuum, as well as the overall reliability of our opacity data.

  The \textit{lower panels} of Fig.~\ref{fig:full_spec} show the smoothed flux spectra, along with the relative difference between \texttt{sunA09}, \texttt{sunA21} simulation and two sets of solar observations. 
  The agreement between (smoothed) synthesised and measured flux is satisfactory above $\sim 450$ nm: Fractional differences between modelling and the \citet{2005MSAIS...8..189K} irradiance are below 3\% in general; The difference between modelling and the \citet{2009GeoRL..36.1101W} spectra is around 3\%, which is close to the maximum uncertainty of the measurement (about 3.5\% in the optical and near-infrared, see Sect.~3 of \citealt{2009GeoRL..36.1101W}). However, notable differences are found below $\sim 400$ nm, where the predicted absolute fluxes are systematically larger than observation by more than 10\% for both solar irradiance datasets.
  As discussed in \citet{2004A&A...417..769A}, \citet{2021A&A...653A..65W} and \citet{2022A&A...657L..11K}, we suspect the discrepancy in the blue end of the spectra is due to missing opacities in the ultraviolet. We also note that several solar model atmospheres (both 1D and 3D) all predict more fluxes than the measured values below the Balmer jump \citep{2018A&A...613A..24K,2021A&A...653A..65W,2022A&A...657L..11K}. Further investigations into the continuum and/or line opacities in the near ultraviolet region might be needed to improve the theoretical-observational consistency of absolute flux in this wavelength range.

\subsection{Continuum centre-to-limb variations} \label{sec:CLV}

\begin{figure*}
\subfigure{
\begin{overpic}[width=0.49\textwidth]{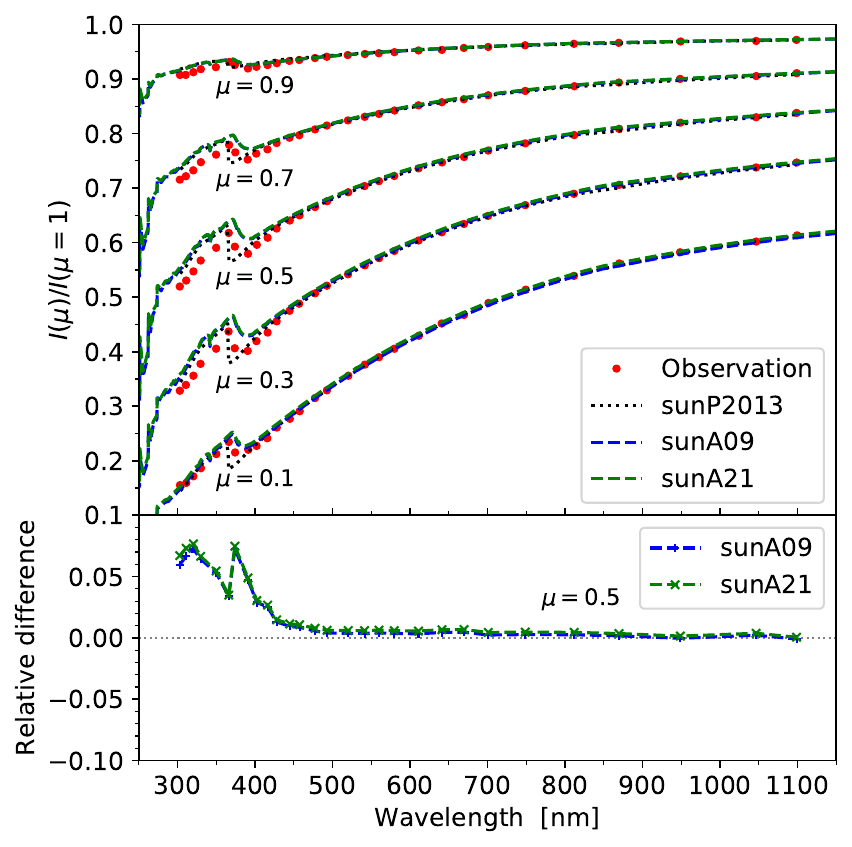}
\end{overpic}
}
\subfigure{
\begin{overpic}[width=0.49\textwidth]{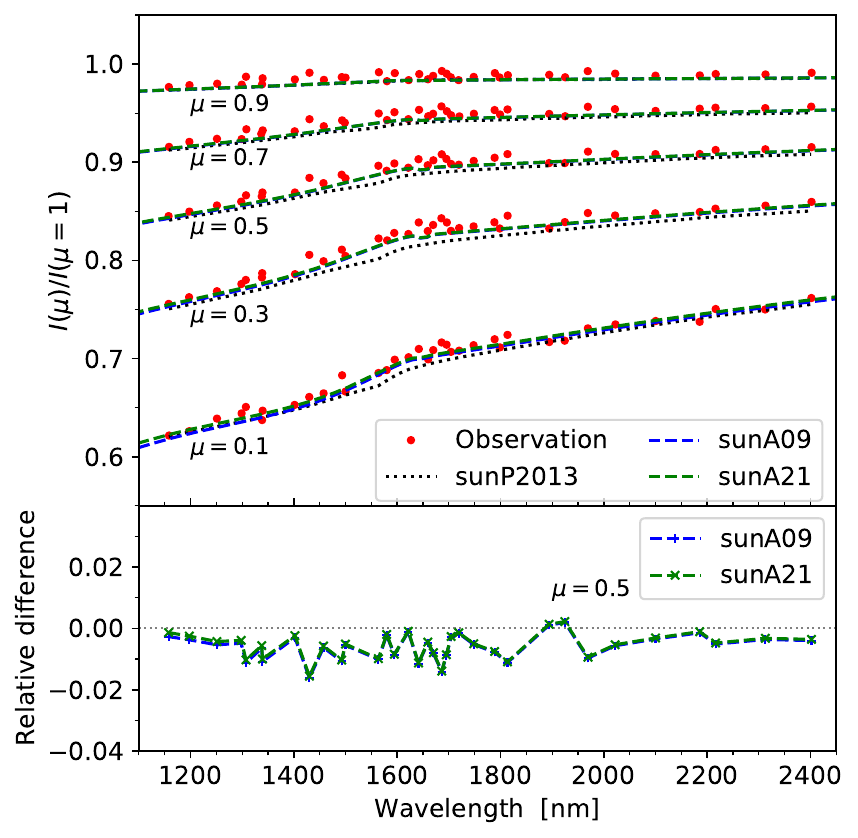}
\end{overpic}
}
\caption{Continuum CLVs at different viewing angles in the optical and near-infrared wavelength range. Observational data (red dots) in the \textit{upper left} and \textit{upper right panel} are taken from \citet{1994SoPh..153...91N} and \citet{1977SoPh...52..179P}, respectively. Blue and green dashed lines are theoretical results of the \texttt{sunA09} and \texttt{sunA21} models, computed using \balder{}. Continuum CLVs predicted by model \texttt{sunP2013} are shown in black dotted lines (cf.~Fig.~3 of P2013). 
Relative differences between the modelled and measured CLVs are demonstrated in the \textit{lower panels}, computed via $[I(\mu)/I(\mu=1)] / [I(\mu)/I(\mu=1)]_{\rm obs} - 1$, where subscript ``obs'' stands for observational data. Only the difference at $\mu = 0.5$ is shown for clarity.
}
\label{fig:comp_CLV}
\end{figure*}

  The magnitude of stellar surface intensity depends on both wavelength and viewing angle. Recall from the Eddington-Barbier approximation that at a given wavelength, the surface intensity at the stellar limb emerges from a smaller optical depth (lower temperature) than the intensity at the disk centre, thereby appearing darker. A detailed understanding of the limb darkening phenomenon is necessary to accurately interpret the light curve of transit exoplanets \citep{2016MNRAS.457.3573E}. Limb-darkening laws are also important for the determination of stellar radii via interferometry: the angular diameter of a star is obtained by fitting the limb-darkened stellar disc model to the visibility curve measured from interferometry \citep{2006A&A...446..635B,2013MNRAS.433.1262W}. Limb darkening can be quantified by the ratio of emergent intensity $I_{\lambda}(\mu) / I_{\lambda}(\mu = 1)$, with $\mu = \cos \theta$, where $\theta$ is the viewing angle relative to disk centre. 
  In this section, we compute the continuum centre-to-limb variation (CLV) for the new solar models and compare them with the corresponding observations. The continuum CLV reflects the temperature stratification in the continuum forming regions, and is therefore often used to check the realism of 3D atmosphere models (e.g., \citealt{2012A&A...539A.121B}, P2013).
  
  Theoretical emergent intensities for different wavelengths and angles were computed using \balder{}. The setup of our radiative transfer calculation is detailed in Sect.~\ref{sec:flux-spec}.
  Between 303.3 nm and 1098.9 nm, our theoretical predictions are compared with the observations of \citet{1994SoPh..153...91N}, where the CLV was measured at multiple continuum wavelengths. Above 1100 nm, the observational data is taken from \citet{1977SoPh...52..179P}. Results from model \texttt{sunP2013} are also included in Fig.~\ref{fig:comp_CLV} for reference.
  
  Fig.~\ref{fig:comp_CLV} reveals a general trend in continuum CLV: it is strong at shorter wavelengths while becoming less pronounced in the near-infrared. For all five angles considered, continuum CLVs predicted by the three 3D model atmospheres are almost indistinguishable at most wavelengths, indicating the temperature gradient between the three models is nearly identical around the optical surface.
  Below 400 nm, predicted CLVs are systematically weaker (i.e.~larger ratios) than observation. This discrepancy is likely associated with difficulties in determining the continuum level in this wavelength region. The near-ultraviolet regime is abundant in spectral lines. CLVs measured at selected wavelengths with finite bandwidths (\citealt{1994SoPh..153...91N} Sect.~2) might contain unaccounted lines which will affect the measured values.
  From 400 nm to about 1300 nm, there is excellent agreement between all synthetic continuum CLVs and observations. At longer wavelengths ($1400 \lesssim \lambda \lesssim 1800$ nm) and closer to the limb ($\mu \leq 0.5$), continuum CLVs predicted by 3D models are systematically stronger than measurements by about 1\%. The discrepancy here is smaller for the \texttt{sunA09} and \texttt{sunA21} models. Overall, the new solar model atmospheres predict continuum CLVs that match well with measurements, performing even better than the solar model of P2013 in the near-infrared.

\subsection{Hydrogen line profiles} \label{sec:Hlines}

\begin{figure*}
\centering
\subfigure{
\includegraphics[width=0.985\textwidth]{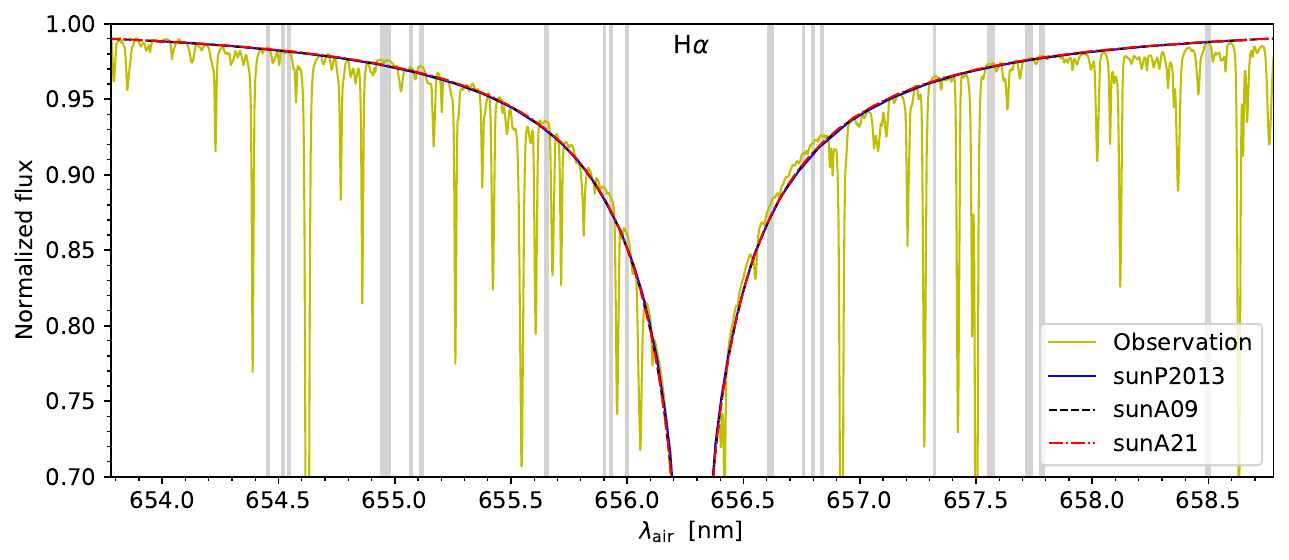}
}

\vspace{-2em}%

\subfigure{
\includegraphics[width=0.99\textwidth]{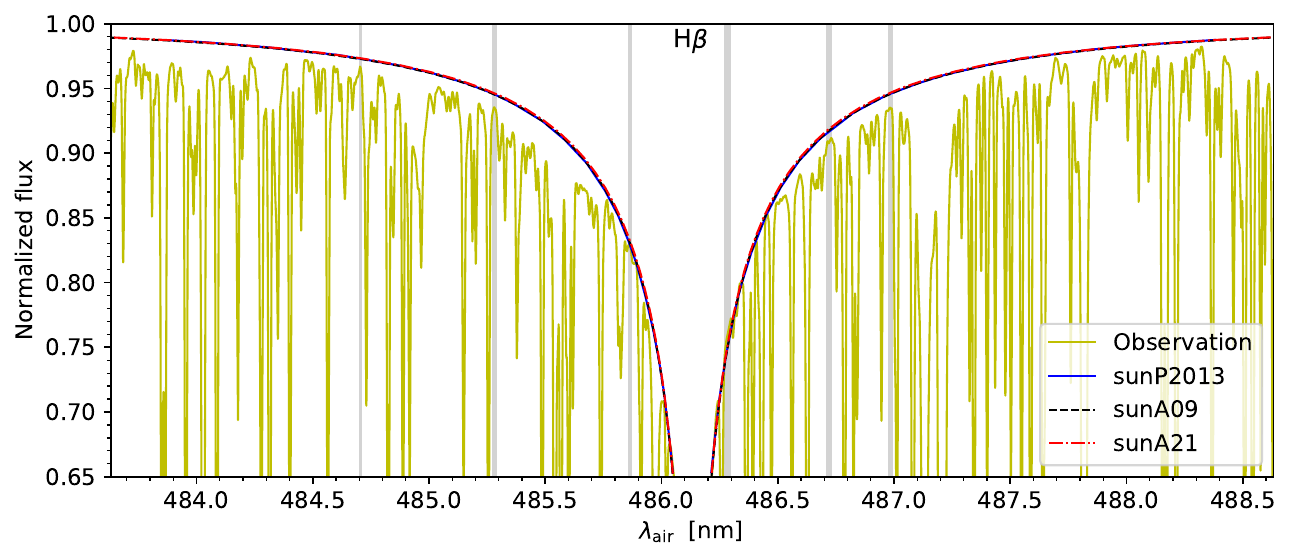}
}

\vspace{-2em}%

\subfigure{
\includegraphics[width=0.99\textwidth]{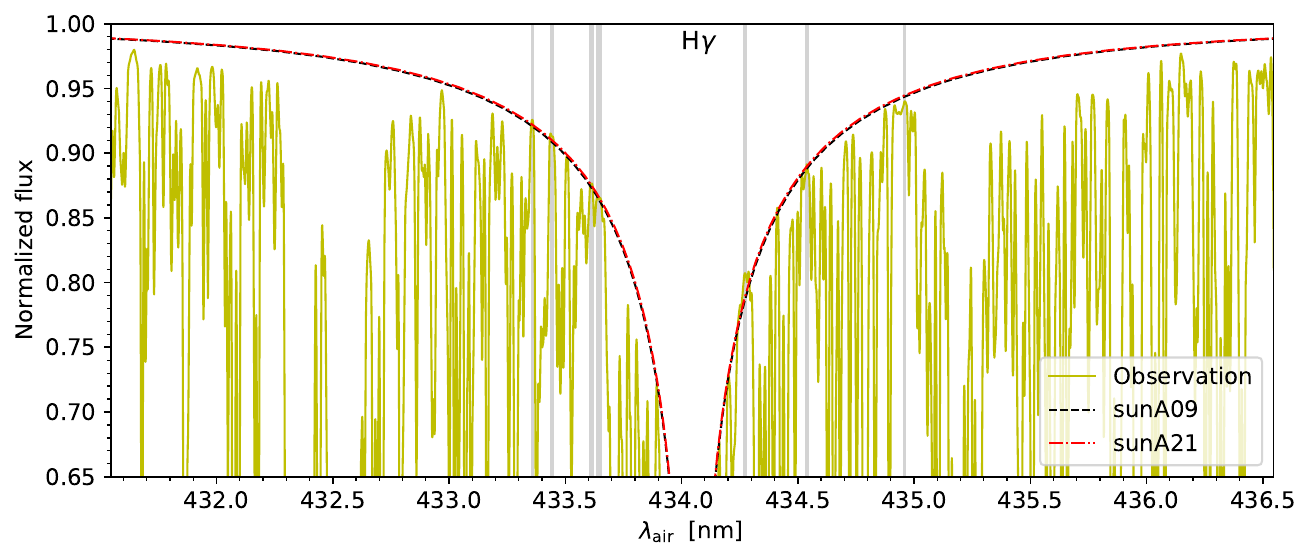}
}

\vspace{-1.5em}%

\caption{Comparison of the synthesised normalised flux profiles with the \citet{2005MSAIS...8..189K} normalised solar flux atlas (yellow lines) for the Balmer series \ce{H\alpha} (\textit{upper panel}), \ce{H\beta} (\textit{middle panel}) and \ce{H\gamma} (\textit{lower panel}), where $\lambda_{\rm air}$ is air wavelength. Black dashed and red dash-dotted lines represent synthesised line profiles computed in non-LTE using \balder{}, based on the \texttt{sunA09} and \texttt{sunA21} models respectively. Theoretical results from the \texttt{sunP2013} model (taken from P2013 Fig.~8) are depicted in blue solid lines. 
Grey vertical bands indicate the ``line masks'' derived in \citet{2018A&A...615A.139A}, which highlights the unblended wavelength sections. 
Only the wings are shown in the figure because line cores, which are formed in the chromosphere, are of no concern to this study.}
\label{fig:HLines}
\end{figure*}

\begin{figure*}
\centering
\subfigure{
\includegraphics[width=0.99\textwidth]{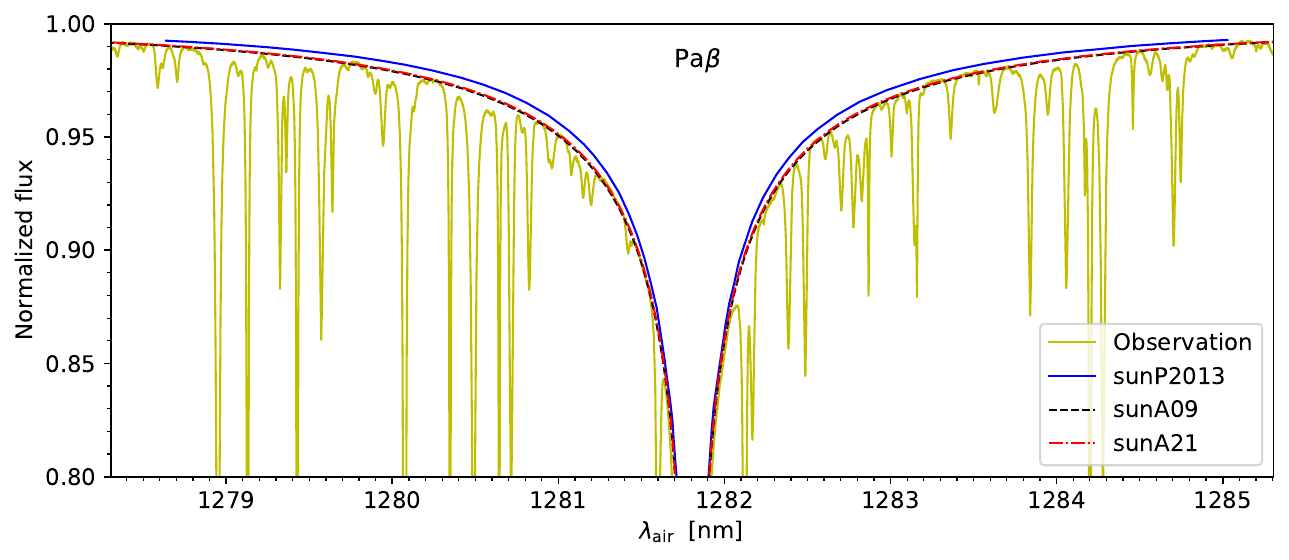}
}

\vspace{-2em}%

\subfigure{
\includegraphics[width=0.99\textwidth]{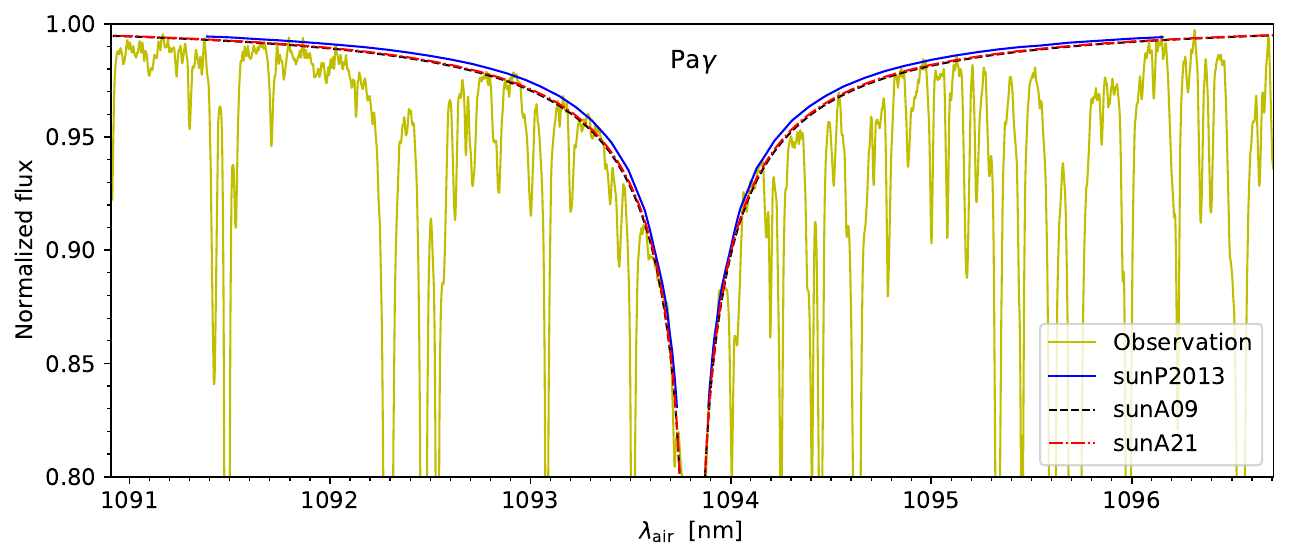}
}

\vspace{-1.5em}%

\caption{Normalised flux profiles for the Paschen lines \ce{Pa\beta} (\textit{upper panel}) and \ce{Pa\gamma} (\textit{lower panel}). All theoretical line profiles are computed in non-LTE. For the \ce{Pa\gamma} line, the measured flux is adopted from \citet{2005MSAIS...8..189K}. The observational data of the \ce{Pa\beta} line, however, is taken from the IAG solar flux atlas \citep{2016A&A...587A..65R} because this wavelength region is not covered by the \citet{2005MSAIS...8..189K} atlas.
}
\label{fig:PaLines}
\end{figure*}

  The spectral lines of hydrogen, in particular the Balmer series, 
  are commonly used to derive the effective temperature for late-type stars owing to their relative insensitivity to the surface gravity and hydrogen abundance (e.g.~\citealt{1993A&A...271..451F,2002A&A...385..951B,2018A&A...615A.139A}). They feature pronounced pressure-broadened wings that form across the lower atmosphere and the surface convection zone ($-2 \lesssim \log\tau_{\rm Ross} \lesssim 1$, Fig.~2 of \citealt{2018A&A...615A.139A}), while the line cores are formed in the chromosphere. As the wings of Balmer lines (especially \ce{H\beta} and \ce{H\gamma}) form in relatively deep layers, their shape is affected by the near-surface convection process. Nevertheless, the wings are largely unaffected by Doppler broadening and shifts due to convective motions, 
  making them suitable probes to the temperature structure of the stellar atmosphere, including the surface convection zone \citep{2009A&A...502L...1L}.
  We follow P2013 in comparing our synthetic spectra to solar observations of the \ce{H\alpha}, \ce{H\beta} and \ce{H\gamma} Balmer lines, as well as the \ce{Pa\beta} and \ce{Pa\gamma} Paschen lines. 

  The synthesised Balmer and Paschen line profiles are presented in Figs.~\ref{fig:HLines} and \ref{fig:PaLines}, together with the normalised solar flux atlases of \citet{2005MSAIS...8..189K} and \cite{2016A&A...587A..65R} for comparison.
  The grey-shaded regions in Fig.~\ref{fig:HLines} are the ``line masks'' for the \ce{H\alpha}, \ce{H\beta} and \ce{H\gamma} lines derived in \citet{2018A&A...615A.139A}. The masks were carefully selected based on theoretical line lists and the observed solar spectrum (cf.~Sect.~4.2.1 of \citealt{2018A&A...615A.139A} for a detailed description) in order to highlight the unblended wavelength sections that reflect only the Balmer lines. This is particularly necessary for a clear identification of the observed \ce{H\beta} and \ce{H\gamma} lines, as their wavelength regions suffer from severe line blending. The line masks were chosen to bracket the wavelength regions that are sensitive to the effective temperature. For Balmer lines, we will inspect how well our theoretical line profiles fit solar observations with the help of these masks.
  
  Results based on the new 3D solar models and \textit{ab initio} non-LTE radiative transfer calculations are in reasonable agreement with measured normalised fluxes for all hydrogen lines considered, indicating the temperature stratification from the surface convection zone and the lower atmosphere of our new models is realistic. 
  Nevertheless, neither the 3D solar model presented in P2013 nor the new models are able to predict line profiles that perfectly match observations for all Balmer and Paschen lines. 
  For the wings of the \ce{H\alpha} line, models \texttt{sunA09} and \texttt{sunA21} predict almost identical profiles as the solar model of P2013, all being smaller than the measured normalised flux. On the other hand, the new solar models give rise to weaker \ce{H\beta} lines particularly in the outer wings ($\gtrsim 0.4$ nm away from the line core).
  Although the wavelength region of the \ce{H\gamma} line is heavily blended, we found reasonable agreement between the synthesised and the measured line profile in the unblended region highlighted by the line masks.
  
  As discussed in P2013, the discrepancy at the \ce{H\beta} wings is not associated with the single line simplification in our line formation calculations, as including the effect of blends in theoretical calculation hardly changes the overall magnitude of the \ce{H\beta} wings. Input physics to the 3D atmosphere model is also unlikely the main cause of this discrepancy, because two sets of 3D solar models with distinct EOS and opacity both failed to perfectly reproduce the observed line profile. In short, the underlying reason why 3D models underestimate the strength of \ce{H\beta} wings is currently unknown.
  
  For Paschen lines, the \texttt{sunA09} and \texttt{sunA21} models predict stronger wings compared with \texttt{sunP2013}. The new solar models perform better in the case of \ce{Pa\beta} line, achieving good agreement with the observed solar spectrum. Conversely, the \ce{Pa\gamma} line computed from the new models deviates a bit further from observations compared to model \texttt{sunP2013}, being slightly stronger than observations especially in the blue wing.
  
  To conclude, hydrogen lines computed based on the new 3D solar models agree with solar observations in general, with some synthetic lines (\ce{H\gamma}, \ce{Pa\beta}) matching observations at a very satisfactory level while others (\ce{H\alpha}, \ce{H\beta}, \ce{Pa\gamma}) demonstrate small deviations. Meanwhile, synthetic lines computed with the \texttt{sunA09} and \texttt{sunA21} models are almost indistinguishable in all cases, implying that different versions of the solar chemical composition have little impact on the hydrogen line profile.

\section{Summary and conclusions}

  In this work, we constructed new 3D solar atmosphere models with the \stagger{} code, using EOS, opacity and solar compositions that are different from previous studies. We adopted the \freeeos{}, an open-source EOS code based on the minimisation of the Helmholtz free energy. Thermodynamic quantities computed via the EOS were tabulated in a format compatible with the \stagger{} code. 
  Monochromatic extinction coefficients were computed from the \blue{} opacity package. In the high temperature region, opacity calculations were based on \freeeos{} for more accurate number densities of all atomic species and better consistency between the EOS and opacity code. 
  Monochromatic extinction coefficients were grouped into 12 different bins to be used in the 3D simulation. Following \citet{2011A&A...528A..32C}, we excluded continuum scattering from the extinction coefficient when calculating the mean intensity weighted mean opacities ($\alpha_J$) in the optically thin part (the \textit{no-scattering-in-streaming-regime} approximation). For each opacity bin, the mean intensity weighted and the Rosseland mean extinction coefficients were merged to obtain the final bin-averaged extinction coefficients. The opacity binning procedure is identical to previous studies \citep{2013A&A...557A..26M,2018MNRAS.475.3369C}.
  It is worth noting that for all models constructed utilising the opacity binning method, the predicted surface flux and effective temperature differ from the monochromatic solution (see \citealt{2023arXiv230603744P} for an in-depth investigation of this problem). For solar models presented in this work, we have carefully optimised the organisation of opacity bins to minimise the error in radiative heating rates and surface flux.
  
  3D solar atmosphere models were constructed with a recent version of the \stagger{} code \citep{2018MNRAS.475.3369C}, based on the aforementioned input physics and the AGSS09 and AAG21 solar abundance. The simulations were properly relaxed and bottom boundary conditions carefully adjusted such that the effective temperature of the model is as close to the reference solar value as possible. The new models employ identical numerical mesh as the \texttt{sunRef} model of \citet{2019ApJ...880...13Z} and AAG21. 
  Being the first time to implement the \freeeos{} and \blue{} opacity to \stagger{} simulations, and noticing that the mean extinction coefficients given by \blue{} show recognisable differences from our previous opacity choice (Figs.~\ref{fig:diff_alphaR} and \ref{fig:diff_alphaP}), it is necessary to test the fidelity of the new models. We first checked the mean structure of the new models by comparing the spatial and temporal averaged quantities with that of the \texttt{sunRef} model.  
  It turns out that the new and the \texttt{sunRef} model agrees well in terms of mean stratification -- for all mean quantities studied, the relative differences are within a few percent in most part of the atmosphere model. Larger discrepancies appear only in the outermost layers of the simulation, where the realism of the model is more uncertain due to other factors such as the magnetic field. 
  We subsequently examined the distribution of disk-centre bolometric intensity and vertical velocity near the optical surface of the new solar models, which reflects the area coverage and relative strength of upflows and downflows at the solar photosphere. Similar good agreements are achieved between the reference and the new solar models. 
  
  Our new solar model atmospheres are not only compared with model \texttt{sunRef} but also validated against various observational constraints. We carried out the radiative transfer post-processing of the new 3D models using the \balder{} code, which employs identical opacity sources as the atmosphere model. The modelled absolute flux spectrum and continuum CLVs were compared with corresponding solar observations as well as results from a well-justified \stagger{} solar model of P2013 (\texttt{sunP2013}). 
  Although different input physics are used in P2013 and this work, our theoretical results match observation well in both tests, performing even better in terms of continuum CLVs in the near-infrared region.
  Moreover, we performed detailed non-LTE line formation calculations for five hydrogen lines with \balder{}. We found that neither of the two new 3D models is able to perfectly reproduce the measured normalised fluxes for all hydrogen lines investigated. Nevertheless, considering the approximations (e.g.~opacity binning) employed in the 3D modelling, and also the performance of 1D solar atmosphere models in this problem (see Fig.~8 of P2013), the wings of the synthetic lines predicted by the new 3D models fit reasonably well with the solar flux atlases, accomplishing similar level of realism as model \texttt{sunP2013}.
  To sum up, the new solar models are able to satisfactorily reproduce observations in all diagnostics, suggesting these \textit{ab initio} simulations predict highly realistic temperature stratification at the top of the convective envelope and the lower atmosphere.
  
  We also notice that the two new models with different solar abundance have very similar structures and predict nearly identical observables in all cases studied. This finding is in line with expectation, as the AGSS09 and AAG21 solar compositions are not drastically different from each other. 
  
  Having passed comprehensive tests against the \texttt{sunRef} model and observations, the validity of the new models as well as their underlying input physics can be confirmed. Therefore, the input physics developed in this work can be applied to the modelling of other stars with confidence. 
  This opens an exciting new path for studying stars with different $\alpha$-element abundance, carbon-enhanced metal-poor stars and population II stars with peculiar chemical compositions, which we were incapable of modelling due to limitations on input physics: Previous studies on key benchmark metal-poor stars were often based on 3D model atmospheres with solar-scaled chemical composition and a fixed value of $\alpha$ enhancement (e.g.~\citealt{2016MNRAS.463.1518A,2022MNRAS.509.1521W}). While this assumption is usually valid, the abundances of certain elements, including carbon, oxygen, and magnesium, have a strong influence on microphysics due to their contribution to either atmospheric opacities or electron density. Neglecting variations in their abundance can lead to undesirable systematic errors \citep{2017A&A...598L..10G,2018A&A...615A.139A}. In the future, we plan to construct 3D model atmospheres with varying $\alpha$ enhancement, carbon-to-oxygen ratio, or abundance patterns tailored for individual cases, as more detailed atmosphere models might improve the accuracy of abundance determination thereby providing insights into the chemical evolution of the Milky Way.

\begin{acknowledgements}
The authors would like to thank the anonymous referee for their careful and thorough report 
that improved the quality of this work. We also thank {\AA}ke Nordlund, Martin Asplund and Regner Trampedach for kindly answering many questions regarding EOS, opacities and 3D modelling. Lionel Bigot's help on opacity binning and intensity distribution is greatly appreciated. We are grateful to Regner Trampedach, {\AA}ke Nordlund, J{\o}rgen Christensen-Dalsgaard, Cis Lagae, Luisa Rodr\'{i}guez D\'{i}az, Tiago Pereira and Karin Lind for reading and commenting on this manuscript. We thank also Sven Wedemeyer and Friedrich Kupka for valuable comments and fruitful discussions. Finally, we thank Alan Irwin for making the \freeeos{} code publicly available.
YZ acknowledges support from the Carlsberg Foundation (grant agreement CF19-0649). AMA gratefully acknowledges support from the Swedish Research Council (VR 2020-03940). Funding for the Stellar Astrophysics Centre is provided by The Danish National Research Foundation (grant agreement no.:DNRF106).
This research was supported by computational resources provided by the Australian Government through the National Computational Infrastructure (NCI) under the National Computational Merit Allocation Scheme and the ANU Merit Allocation Scheme (project y89). This work was supported by the Ministry of Education, Youth and Sports of the Czech Republic through the e-INFRA CZ (ID:90254).
\end{acknowledgements}


\bibliographystyle{aa} 
\bibliography{References.bib}



\begin{appendix}

\section{Comparison between MHD and \freeeos} \label{sec:compare-EOS}

\begin{figure*}
\subfigure{
\begin{overpic}[width=0.49\textwidth]{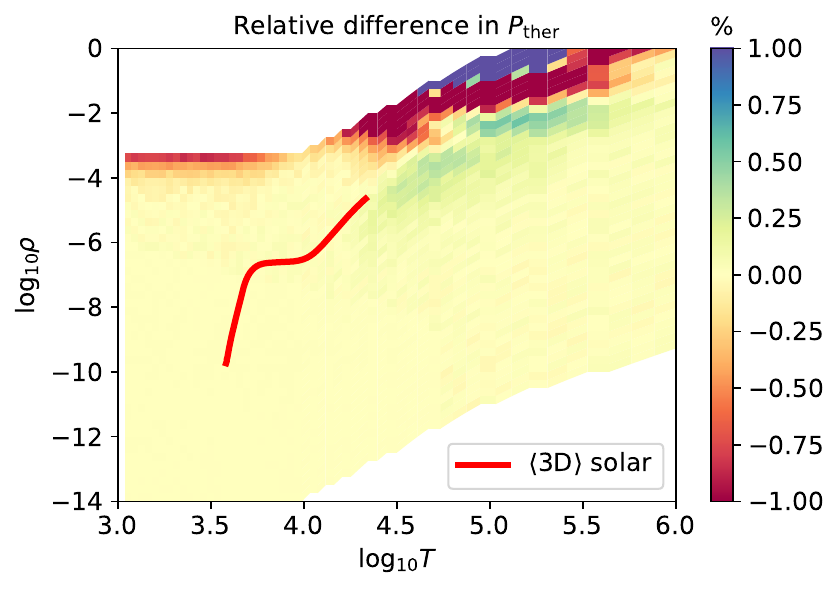}
\end{overpic}
}
\subfigure{
\begin{overpic}[width=0.49\textwidth]{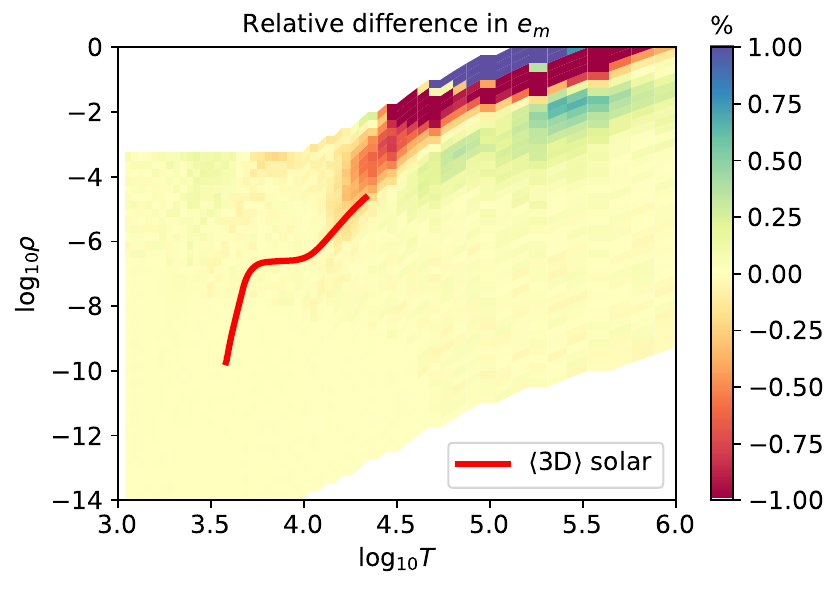}
\end{overpic}
}
\caption{Comparison of thermodynamic quantities between the MHD and \freeeos{}.
\textit{Left panel:} Relative differences between the MHD and \freeeos{} in thermal pressure, computed via $(P_{\rm ther,Free} - P_{\rm ther,MHD}) / P_{\rm ther,MHD}$. Data used in the comparison is generated directly from the corresponding EOS code and based on identical $(\rho,T)$ values, therefore no interpolation is involved in the comparison. The mean structure of our reference 3D solar model atmosphere is over-plotted in the red line, indicating the area of most interest. Both EOSs are computed assuming the same chemical composition, which includes 15 elements with the AGSS09 solar abundance value. Densities and temperatures are in cgs unit.
\textit{Right panel:} Relative differences between the MHD and \freeeos{} in internal energy per mass.
}
\label{fig:diff_EOSrhoT}
\end{figure*}

  To validate the results from \freeeos{}, we compare key EOS outputs with the corresponding quantities from the (modified version of the) MHD EOS table, which was our adopted EOS in previous simulations (e.g.~\texttt{sunRef} mentioned in Sect.~\ref{sec:solar-model}, the solar model presented in P2013 and the \stagger{}-grid models of \citealt{2013A&A...557A..26M}). Fig.~\ref{fig:diff_EOSrhoT} shows the comparisons of two of the output parameters -- thermal pressure and internal energy per mass. Our comparison adequately covers the parameter space researchable by surface convection simulations of low-mass stars. For reference, the mean structure of the horizontal- and time-averaged 3D solar model is plotted in red in Fig.~\ref{fig:diff_EOSrhoT} to indicate the approximate area of most interest. 
  
  As shown in Fig.~\ref{fig:diff_EOSrhoT}, the agreement between MHD and \freeeos{} is overall excellent, with the relative difference in key thermodynamic quantities far less than 1\% at most $(\rho,T)$ range considered. This is reassuring and indicates that our \freeeos{} results are reasonable. Meanwhile, differences are found in both comparisons in the high-temperature ($\log(T / {\rm [K]}) \gtrsim 4.5$) and high-density ($\log(\rho / {\rm [g/cm^3]}) \gtrsim -5$) area. As the difference appears in both comparisons at the same region, it is likely that the area is subject to a difference in the general treatment of some physical aspect between the two EOS codes rather than numerical issues. 
  \citet{1988ApJ...331..794H} implemented a $\tau$-correction (not to be confused with optical depth) in the MHD EOS in order to limit the otherwise diverging, first-order, Debye-H\"uckel term of the Coulomb pressure. In comparison with the OPAL EOS, \citet{2006ApJ...646..560T} found the suppression by $\tau$ to be too strong and result in much better agreement between the two EOSs if left out completely. The \freeeos{} was written and originally tested with respect to a version of the MHD EOS (and OPAL EOS) including the $\tau$-correction. 
  Moreover, \citet{2006ApJ...646..560T} showed the $\tau$-correction to be especially relevant in areas of higher densities. As the observed differences lie at high densities, this suggests that the differences are, at least in part, due to the $\tau$-correction still being implemented in the \freeeos{}. However, modifying the \freeeos{} code and changing the $\tau$-correction implementation is beyond the scope of this work.

\section{The effect of EOS on the mean temperature stratification} \label{sec:compare-T}

\begin{figure}
\begin{overpic}[width=\columnwidth]{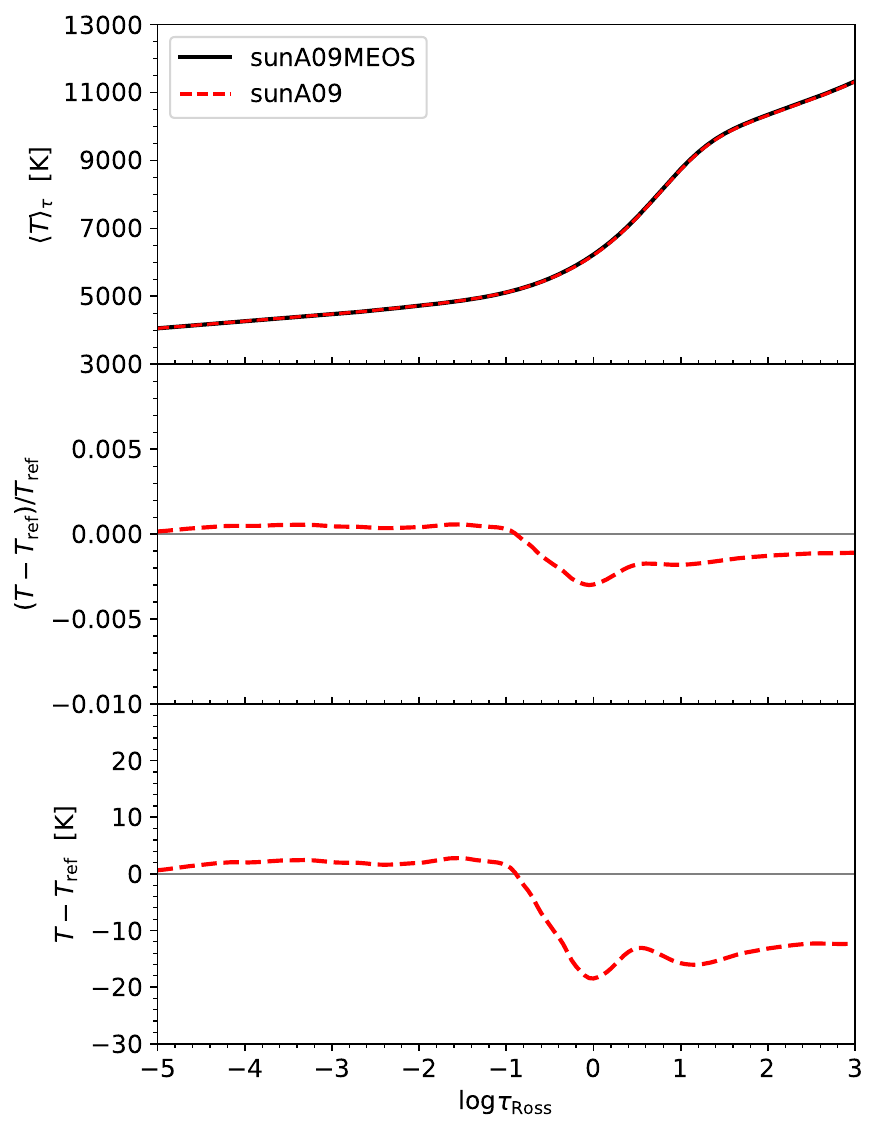}
\end{overpic}
\caption{The $\tau_{\rm Ross}$- and time-averaged temperature as a function of Rosseland optical depth for \texttt{sunA09} (identical to the red dashed line in the \textit{upper right panel} of Fig.~\ref{fig:comp_T}) and a model constructed with the MHD EOS (black solid line labelled \texttt{sunA09MEOS}). The two solar models are identical in all aspects except for employing different EOSs. In the \textit{middle} and \textit{bottom panel}, the model based on MHD EOS was used as a reference to compute relative and absolute differences.
}
\label{fig:comp_sunA09_EOS_T}
\end{figure}

  To study how different EOSs affect the mean temperature structure of the 3D solar model atmosphere, we compare two 3D models based on \freeeos{} (\texttt{sunA09}) and MHD EOS but identical in all other aspects such as opacity data and numerical setup. For both models, their initial simulation snapshot was constructed from the same density and pressure datacube of a relaxed simulation, with respective EOS used to derive initial internal energy and temperature (see also \citealt{2015angeo-Vitas} Sect.~3). The initial simulation snapshot of the two models underwent identical relaxation processes, after which they were evolved for the same time duration for fair comparisons.

  Spatial and temporal averaged temperature for the two models, as well as their realtive and absolute difference, is demonstrated in Fig.~\ref{fig:comp_sunA09_EOS_T}. Using the MHD or \freeeos{} could lead to about 15 K temperature difference around the optical surface and in the near-surface convective region. The impact of EOS on atmospherical temperature structure is smaller than 5 K for $\tau_{\rm Ross} \lesssim 0.1$. This is in line with the conclusion of \citet{2015angeo-Vitas}, who investigated the effect of EOS on the temperature structure for 2D \texttt{MURaM} solar near-surface convection simulations and found that different EOSs cause less than 10 K temperature difference in the atmosphere.

\section{Comparing mean opacities} \label{sec:compare-op}

\begin{figure*}
\centering
\includegraphics[width=0.49\textwidth]{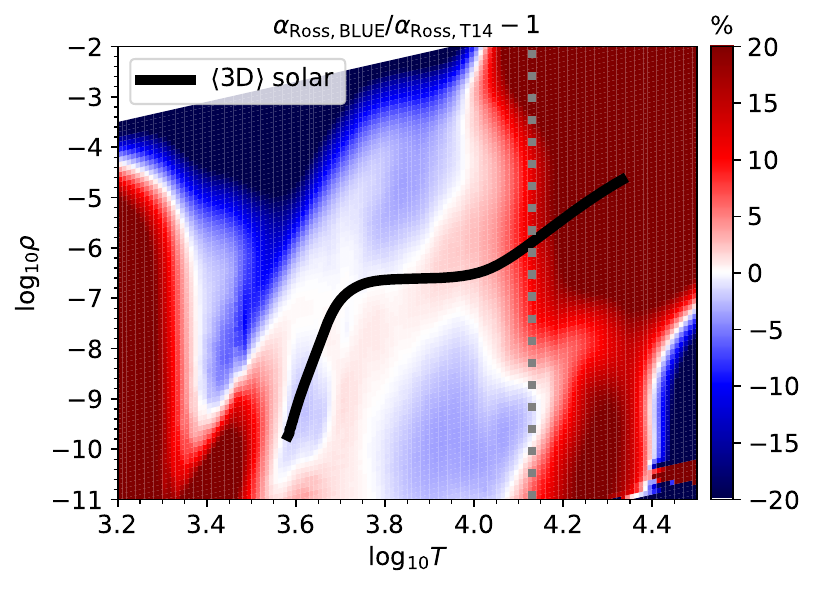}
\includegraphics[width=0.49\textwidth]{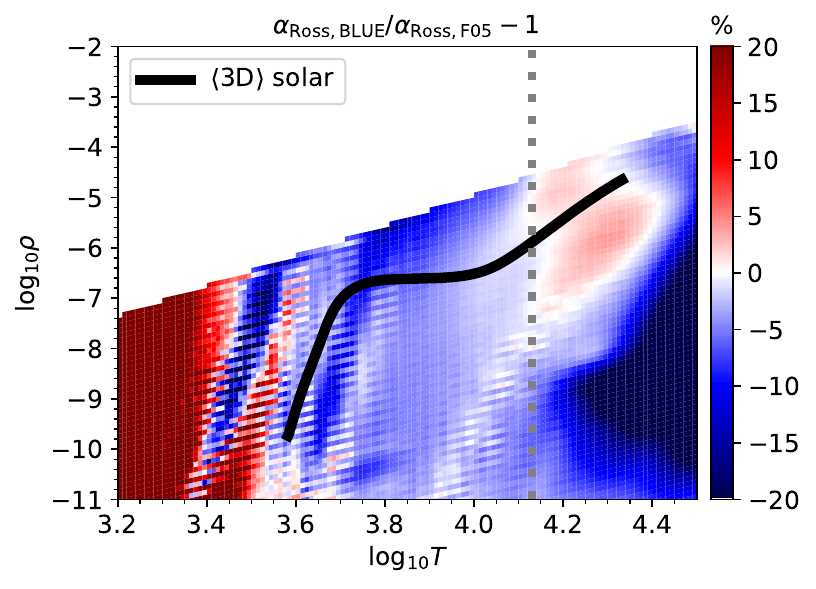}
\caption{Comparison of Rosseland mean extinction coefficients.
\textit{Left panel:} Relative differences between Rosseland mean extinction coefficients from \blue{} and that previously adopted in the \stagger{} code \citep{2010A&A...517A..49H,2014MNRAS.442..805T}, computed via $(\alpha_{\rm Ross,BLUE} - \alpha_{\rm Ross,T14}) / \alpha_{\rm Ross,T14}$. 
\blue{} extinction coefficients are calculated with the internal EOS in \blue{} (\freeeos{}) when $\log T \leq 3.9$ ($\log T \geq 4.1$). Results from the two different EOSs are merged in the region where $3.9 < \log T < 4.1$ (cf.~Sect.~\ref{sec:op-calc}).
The $(\rho,T)$ distribution of the horizontal- and time-averaged reference solar model atmosphere is over-plotted in thick black line. The grey dotted vertical line marks crudely the temperature limit above which the radiative transfer equation is not solved in the reference solar simulation. Densities and temperatures are in cgs unit.
\textit{Right panel:} Relative differences between Rosseland mean extinction coefficients from \blue{} and \citet{2005ApJ...623..585F} for the AGSS09 solar composition. The white area are $(\rho,T)$ combinations not covered by the \citet{2005ApJ...623..585F} opacity table.
}
\label{fig:diff_alphaR}
\end{figure*}

\begin{figure}
\begin{overpic}[width=\columnwidth]{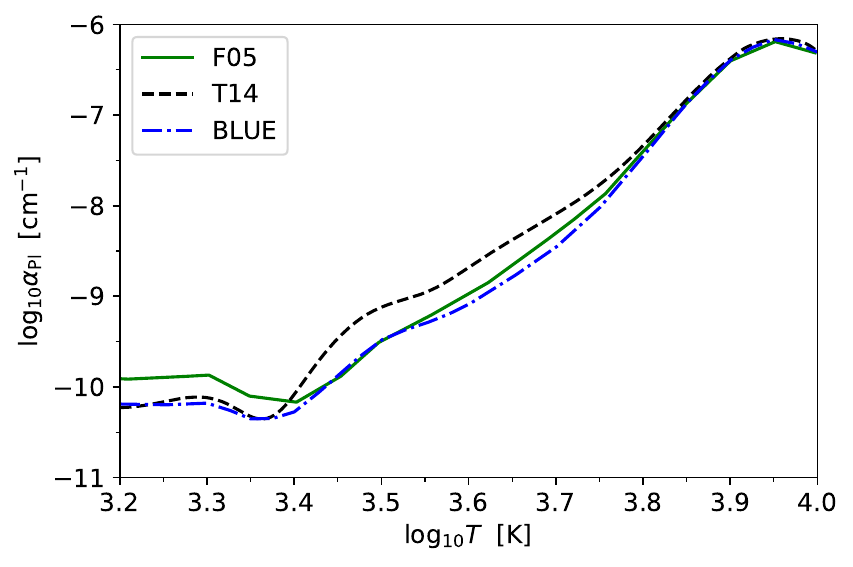}
\end{overpic}
\caption{Comparison between the Planck mean extinction coefficient at $\log(\rho / \rm{[g/cm^3]}) = -10$ from \blue{} and other sources. The green solid line and black dashed line represents Planck mean extinction coefficients from \citet[see also their Fig.~12]{2005ApJ...623..585F} and \citet[the one previously used in the \stagger{} code]{2014MNRAS.442..805T}, respectively. The blue dash-dotted line is $\alpha_{\rm Pl}$ computed using \blue{}, with $250\,000$ wavelength points between 50 nm and 50 $\rm \mu m$. All calculations are based on the AGSS09 solar chemical composition. 
}
\label{fig:diff_alphaP}
\end{figure}

  Relative differences between the Rosseland mean extinction coefficient from \blue{} and that previously adopted in the \stagger{} code are shown in the \textit{left panel} of Fig.~\ref{fig:diff_alphaR} for the AGSS09 solar abundance. The latter adopts a comprehensive source of continuum opacities as elaborated in \citet{2010A&A...517A..49H} and \citet{2014MNRAS.442..805T}, while line opacities are taken from the MARCS package \citep{2008A&A...486..951G,2008PhST..133a4003P}.
  The relative difference in Rosseland mean extinction coefficient is generally below 10\% in our area of interest. However, the difference becomes large when $\log(T / {\rm [K]}) \lesssim 3.3$ and $\log T \gtrsim 4.2$. The underlying reason for this disagreement is not clear. As the two types of opacities are calculated based on different atomic data, partition function, continuum opacity sources etc.~and each factor could lead to the observed difference, investigating the source of the disagreement is beyond the scope of this work. Nevertheless, the areas where a large difference in $\alpha_{\rm Ross}$ is seen have little impact on our modelling at least for the solar case, because (1) temperatures in the simulation domain hardly drop below 2000 K (2) the \stagger{} code no longer solves the radiative transfer equation at regions with very high temperature (that is, regions located well below the photosphere), because the diffusion limit of heat transfer is a good approximation there.
  We also compared Rosseland mean extinction coefficients given by \blue{} with the corresponding \citet{2005ApJ...623..585F} table and found better agreement at the high-temperature, high-density regime (see the \textit{right panel} of Fig.~\ref{fig:diff_alphaR}).
  
  Because the Rosseland mean extinction coefficient is obtained by integrating $\alpha_{{\rm tot},\lambda}^{-1}$, it mainly reflects the contribution from low-value continuum opacities. The Planck mean extinction coefficient, on the other hand, is dominated by strong line opacities. In Fig.~\ref{fig:diff_alphaP}, we compare our Planck mean extinction coefficients at a given density with corresponding data from \citet{2005ApJ...623..585F} and \citet{2014MNRAS.442..805T}. Below $\log T \sim 3.4$, the agreement between \blue{} and \citet{2014MNRAS.442..805T} is good, but they are both smaller than the \citet{2005ApJ...623..585F} result. The cause of this disagreement between \blue{} and \citet{2005ApJ...623..585F} is not clear, but it might be due to different $\rm H_2 O$ and TiO line lists adopted in \blue{} and \citet[see their Fig.~4]{2005ApJ...623..585F}. Above $\log T \sim 3.4$, Planck mean extinction coefficients predicted by \blue{} agree well with \citet{2005ApJ...623..585F}. However, the disagreement between \blue{} and \citet{2014MNRAS.442..805T} within $3.4 \lesssim \log T \lesssim 3.7$ is not understood.

\end{appendix}

\end{document}